\documentclass[journal]{IEEEtran}

\usepackage{amsfonts}
\usepackage{cite,graphicx,amsmath,amsthm,amssymb}
\usepackage{subfigure}
\usepackage{fancyhdr}
\usepackage{dsfont}
\usepackage{array,color}
\usepackage{bm}
\usepackage{float}
\usepackage{algorithm}
\usepackage{algpseudocode}
\usepackage{multirow}
\usepackage{booktabs}
\usepackage{multirow}
\usepackage{makecell}
\usepackage{colortbl}
\usepackage{graphicx}
\usepackage{stfloats}
\usepackage{subfigure} 
\usepackage{cite}

\allowdisplaybreaks[4]

\newtheorem{theorem}{Theorem}

\newtheorem{lemma}{Lemma}

\newtheorem{proposition}{Proposition}

\newtheorem{corollary}{Corollary}

\newtheorem{property}{Property}

\newtheorem{remark}{Remark}

\newtheorem{claim}{Claim}

\begin{document}

	\title{Reconfigurable Intelligent Surface Assisted OFDM Relaying: Subcarrier Matching with Balanced SNR}
	
	\author{ 		
		Tong Zhang, \textit{Member, IEEE}, Shuai Wang, \textit{Member, IEEE}, Yufan Zhuang, 
		Changsheng You, \textit{Member, IEEE}, Miaowen Wen, \textit{Senior Member, IEEE},
		and Yik-Chung Wu,	\textit{Senior Member, IEEE}  
		\thanks{			
			T. Zhang, Y. Zhuang, and C. You are with the Department of Electrical and Electronic Engineering, Southern University of Science and Technology, Shenzhen, China (e-mail:  \{zhangt7, wangs3, 11912327, youcs\}@sustech.edu.cn).}
		\thanks{S. Wang is with Shenzhen Institute of Advanced Technology, Chinese Academy of Sciences, Shenzhen, China (e-mail: s.wang@siat.ac.cn).}
		\thanks{M.~Wen is with the School of Electronic and Information Engineering, South China University of Technology, Guangzhou 510641, China (e-mail: eemwwen@scut.edu.cn).} 
		\thanks{Y.~Wu is with the Department of Electrical and Electronic Engineering, The University of Hong Kong, Hong Kong (e-mail:
			ycwu@eee.hku.hk).} 	
		\thanks{Corresponding author: Miaowen Wen.}				 
	}

	\maketitle

	\begin{abstract}
		This paper considers a reconfigurable intelligent surface (RIS) aided orthogonal frequency division multiplexing (OFDM) relaying system, and investigates the joint design of RIS passive beamforming and subcarrier matching under two cases, where Case-I ignores the source-RIS-destination signal, while Case-II explores this signal for rate enhancement. We formulate a mixed-integer nonlinear programming (MINIP) problem to maximize the sum achievable rate of all subcarriers by jointly optimizing the passive beamforming and subcarrier matching. To solve this problem, we first develop a branch-and-bound (BnB)-based alternating optimization algorithm for attaining a near-optimal solution.  Then, a low-complexity difference-of-convex penalty-based algorithm and learning-to-optimize approach are also proposed. Finally, simulation results demonstrate that the RIS-assisted OFDM relaying system achieves a substantial achievable rate gain as compared to that without RIS since RIS recasts the subcarrier matching and balances the signal-to-noise ratio (SNR) among different subcarrier pairs.  
	\end{abstract}
	

	\section{Introduction}
	
	The reconfigurable intelligent surface (RIS) has recently emerged as
	a promising solution to enhance the performance of wireless communications via smartly reconfiguring the wireless propagation environment \cite{2003,Qingqing}. Specifically, the RIS is a programmable electromagnetic surface, consisting of a massive number of reflecting elements, each capable of tuning the amplitude and/or phase of the reflected signal. As such, the wireless channel can be dynamically programmed by RIS to improve communication performance. The RIS has been studied in various communication scenarios, including edge intelligence \cite{2002,2007,32}, wireless positioning \cite{00,01,02}, unmanned aerial vehicle (UAV) communications  \cite{61,2010,ChangshengUAV,Abdalla}, orthogonal frequency division multiplexing (OFDM) systems \cite{60,600,2006}, simultaneous transmitting and reflecting networks \cite{Yi,2000,2005}.

	In particular, the communication performance comparison between the active relay and RIS has been studied in \cite{301,299,300,302,303}. Specifically, the key differences and similarities between RIS and decode-and-forward (DF) relay were discussed in \cite{301}. For brevity, we henceforth use ``relay" to represent ``DF relay". Besides, it was shown in \cite{299} and \cite{300} that RIS can achieve higher energy and spectral efficiency than active relays due to its passive reflection and full-duplex operation mode. Moreover, the selection of Poisson point distributed RIS and relay networks was studied in \cite{303}, where RIS is shown to achieve better performance in the near-field, in terms of both outage probability and energy efficiency.

	To further improve the communication performance in relaying/RIS networks, the recent works in \cite{198,199,200,201,202,203,204,205} have studied the integration of RIS and relay in wireless networks, rather than unilaterally considering one of them. To be specific, a RIS-assisted relaying system was studied in \cite{198}, where tight upper bounds on the ergodic capacity are obtained under different channel environments. In \cite{199}, the authors showed that  a RIS-assisted full-duplex relaying can achieve significant performance gains over
	half-duplex relay and the
	RIS-only systems. In \cite{200}, the coverage extension by RIS-assisted relaying was investigated to establish blockage-free communication links. In \cite{201}, the time allocation and RIS passive beamforming were jointly optimized for achievable rate maximization in the RIS-assisted relaying.  Moreover, the authors in \cite{202} studied the deployment strategy in the RIS-assisted relaying system and showed that the multi-RIS relaying system attains a higher capacity order. In \cite{203}, a multi-RIS relaying system in Nakagami-$m$ fading channels was investigated, where closed-form expressions are derived for the upper bound
	on the ergodic capacity. In \cite{204}, a deep reinforcement learning algorithm was proposed to design the relay selection scheme in RIS-assisted cooperative networks, which is shown to achieve significant performance gains as compared to random relay
	selection and random passive beamforming designs. Furthermore, a multi-agent deep reinforcement learning-based algorithm was proposed for devising the buffer-aided relay selection scheme for integrated relay and RIS secure networks in the presence of an eavesdropper introduced in \cite{205}.   
	
	\begin{figure}[t]
		\centering
		\includegraphics[width=3.2in]{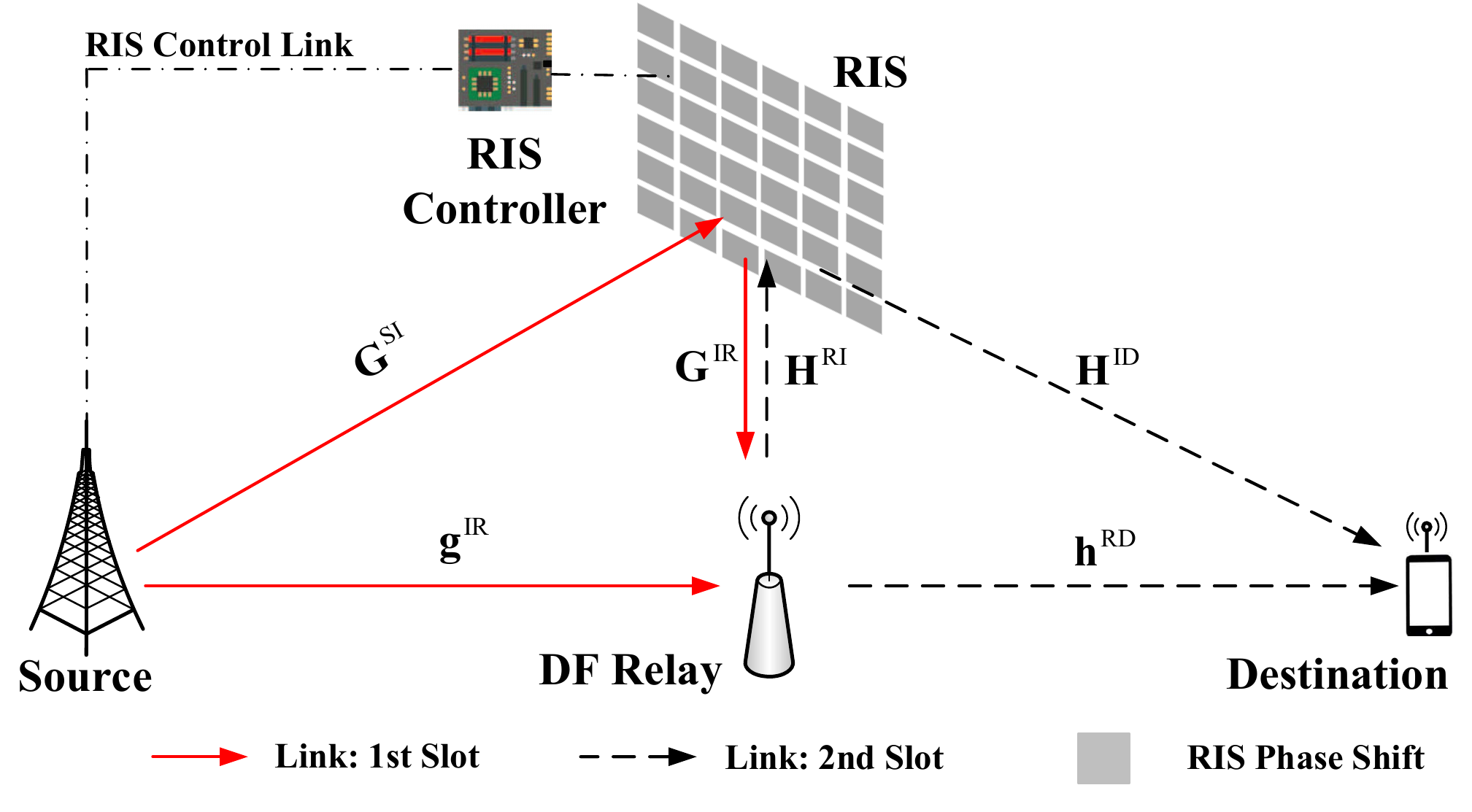}
		\caption{RIS-assisted OFDM relaying for Case-I.} \label{F2}
	\end{figure}

	It is worth noting that the above work mainly focused on the single-carrier relaying systems aided by RIS. However, for future B5G and 6G cellular networks, multi-carrier communication based on OFDM is still mainstream for boosting system capacity. Specifically, compared with single-carrier relaying, OFDM relaying brings about additional degrees-of-freedom for the relay to pair the incoming and outgoing subcarriers according to their signal-to-noise ratios (SNRs), therefore substantially improving the relaying performance \cite{500,501}. Conventionally, the optimal subcarrier matching for maximizing the sum-SNR over all subcarrier pairs is achieved by the best-to-best (BTB) scheme, i.e., the best subcarrier of the first hop with the highest SNR is paired with the best subcarrier of the second hop, and so on. On the other hand, to minimize the bit-error-rate (BER), the optimal subcarrier matching scheme is instead the best-to-worst (BTW) matching, i.e., the best subcarrier of the first hop is matched with the worst subcarrier of the second hop, and so forth. However, for RIS-aided OFDM relaying systems, the effective channel gains over different subcarriers hinge on the design of RIS passive beamforming. This leads to a new subcarrier matching design in the RIS-assisted OFDM relaying system as compared to that without RIS. This, however, has not been well studied in the existing literature and hence motivates the current work.

\begin{figure}[t]
		\centering
		\includegraphics[width=3.2in]{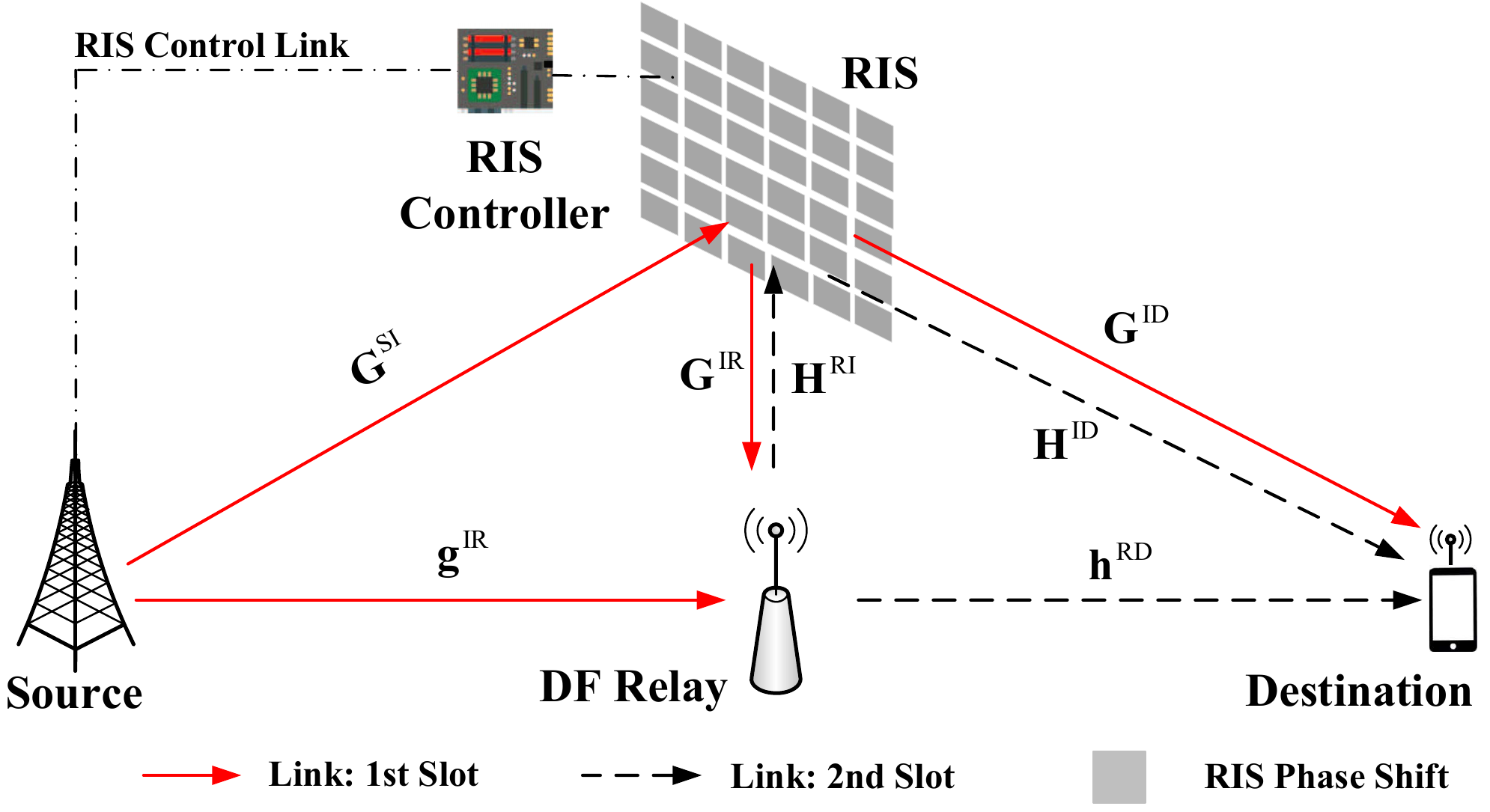}
		\caption{RIS-assisted OFDM relaying for Case-II.} \label{F22}
	\end{figure}

	
	In this paper, we investigate the joint RIS passive beamforming and subcarrier matching in RIS-assisted OFDM relaying networks, where a RIS is deployed to assist the data relaying at a DF relay from a single-antenna source to a single-antenna destination.   We consider two RIS relaying systems, where Case-I, shown in Fig. \ref{F2},  ignores the source-RIS-destination signal in the first hop, while Case-II, shown in Fig. \ref{F22}, explores the source-RIS-destination signal in the first hop.   For these two cases, we formulate and solve a mixed-integer nonlinear programming (MINIP) problem, by jointly designing the RIS passive beamforming and subcarrier matching. The contributions are summarized as follows:
	
	\begin{itemize}	
		\item We first devise a branch-and-bound (BnB)-based alternating optimization algorithm for solving the formulated MINIP problem. This algorithm divides the MINIP problem into the subcarrier matching subproblem and RIS passive beamforming subproblem, where the subcarrier matching and the RIS passive beamforming are solved via BnB and semidefinite relaxation (SDR), respectively. 
		
		\item To reduce the computational complexity of BnB, we further design a difference-of-convex penalty-based algorithm to suboptimally solve the subcarrier matching problem, where the binary constraint is transformed into its continuous counterpart with a penalty on the relaxation. Moreover, to further reduce the complexity, we utilize the learning-to-optimize approach, imitating the difference-of-convex penalty-based algorithm via deep learning. 
		
		\item Simulation results show that the achievable rates of the proposed algorithms in Case-II are higher than that in Case-I when there is a blockage between source-relay and relay-destination links. The achievable rates of the proposed algorithms with RIS are much higher than that without RIS, especially when the number of reflecting elements is large. In contrast to the conventional OFDM relaying system with different SNRs in each paired subcarrier, the RIS-assisted system achieves balanced SNRs over different subcarrier pairs. The complexity of the proposed algorithms is also verified by simulations. 
	\end{itemize}

	\textit{Notations}: The scalar, vector, and matrix are denoted by $\phi,{\bm{\phi}}$, and ${\bm{\Phi}}$, respectively.  The conjugate-transpose operator is denoted by $(\cdot)'$. The identity matrix with $n$ dimensions is denoted by $\textbf{I}_n$. The all-zeros matrix with $n$ rows and $m$ columns is denoted by $\textbf{0}_{n \times m}$. The $n$-dimensional complex Gaussian distribution with zero mean and covariance matrix $\textbf{I}_n$ is denoted by $\mathcal{CN}(0,\textbf{I}_n)$. The Euclidean norm is denoted by $\| \cdot\|$. The assemble of variables $x_{p,q},\forall p,q$ is denoted by $\{x_{p,q}\}$. For matrix ${\bm{\Phi}}$, $\text{Tr}\{{\bm{\Phi}}\}$ denotes its trace, while ${\bm{\Phi}} \succeq \textbf{0}$ means that ${\bm{\Phi}}$ is positive semidefinite. The element of the $n^\text{th}$ row and $m^\text{th}$ column of matrix ${\bm{\Phi}}$ is denoted by ${\bm{\Phi}}(n,m)$. The discrete Fourier transform (DFT) function is denoted by $\mathcal{F}\{\cdot\}$. The convolution operator is denoted by $*$. 
	
	\textit{Organization}: The remaining of this paper is organized as follows: The system model is defined in Section II. The problem formulations are introduced in Section III. We develop the BnB algorithm in Section IV, the difference-of-convex penalty-based algorithm in Section V, and the learning-to-optimize approach in Section VI. The simulation results are presented in Section VII. Finally, conclusions are drawn in Section VIII.

	\section{System Model}
	
	We consider a RIS-assisted OFDM relaying system, which consists of a source, a half-duplex  relay, a RIS, and a destination\footnote{The RIS design in more general systems with two or more RISs and relays faces new challenges, such as multi-IRS reflection, RIS
		deployment and association/selection, which thus is left for our future work \cite{you2022deploy}.}. The bandwidth of each subcarrier in the OFDM system is denoted by $\Delta$. Each subcarrier has independent channel coefficients.  There are a total of $N$ subcarriers for the OFDM system. The source, relay, and destination are equipped with a single antenna. We assume that the distance between source and destination is sufficiently large such that there is no direct link between them.  The relay operates in a half-duplex mode with a time division and DF protocol. For data transmission, the time is divided into two slots, where the duration per time slot is denoted by $T$. In particular, in the first time slot, the source transmits the signal, and the relay receives and decodes the signal. In the second time slot, the relay forwards the decoded signal to the destination. The OFDM  system pairs one subcarrier in the first hop with one subcarrier in the second hop for relaying, where the number of subcarrier pairs is up to $N$. Furthermore, the RIS has $M$ reflecting elements and is connected to a controller for adjusting the passive beamforming. The frequency band in the first hop will be reused in the second hop. Thus, subcarrier assignment is needed for the overall performance of relaying.

	\subsection{Case-I: No RIS-Destination Link in Time Slot One}

	As shown in Fig. \ref{F2}, Case I refers to the scenario where the impact of the RIS-destination link is ignored in the first time slot, which is assumed in most existing works (see, e.g., \cite{198,199,200,201,202,203,204,205}). We assume quasi-static block-fading channels, where all the channels in the same time slot remain approximately constant. In the following, we first introduce the time-domain one-tap channel, and then the wideband OFDM. Denote the time-domain channel\footnote{The RIS channels can be acquired by the customized RIS channel estimation methods (see, e.g, \cite{600,601})}   for the source-relay link in time slot $t=1,2$ by $\widetilde{g}^{\text{SR}}[t] \in \mathbb{C}^1$. Denote the time-domain channel vector for source-RIS link in time slot $t$ by  $\widetilde{\textbf{h}}^{\text{SI}}[t] \in \mathbb{C}^{M \times 1}$. 
	Denote the time-domain channel vector for the RIS-relay link in time slot $t$ by $\widetilde{\textbf{h}}^{\text{IR}}[t] \in \mathbb{C}^{M \times 1}$. Denote the time-domain channel vector for the RIS-destination link  in time slot $t$ by $\widetilde{\textbf{h}}^{\text{ID}}[t] \in \mathbb{C}^{M \times 1}$. Denote the time-domain channel vector for the relay-RIS link in time slot $t$ by $\widetilde{\textbf{h}}^{\text{RI}}[t] \in \mathbb{C}^{M \times 1}$. Denote the time-domain channel for the relay-destination link in time slot $t$ by $\widetilde{g}^{\text{RD}}[t] \in \mathbb{C}^1$.  According to the time division and DF protocol, the source transmits data symbols in the first time slot, where the narrowband transmit signal in time slot $1$ is given by 
	\begin{equation}
		x[1] = s. \label{A1}
	\end{equation}
	Thus, in the time slot $1$, the narrowband received signals at the relay, after the reflection of RIS, can be expressed as
	\begin{equation}
		r^R[1] = \widetilde{g}^{\text{SR}}[1] s + (\widetilde{\textbf{h}}^{\text{IR}}[1])'{\bf{\Phi}}[1]\widetilde{\textbf{h}}^{\text{SI}}[1] s + n^R, \label{A2}
	\end{equation}  
	where the additive white Gaussian noise (AWGN) at the relay is denoted by $n^R$, and the RIS passive beamforming matrix in the first time slot is denoted by ${\bf{\Phi}}_1 = \text{diag}\{\phi_1[1], \phi_2[1], \cdots, \phi_M[1]\}$ with $\phi_m[1] = e^{j\theta_m[1]},\theta_m[1] \in [0,2\pi],\forall m = 1,2,\cdots,M$.  Afterward, in narrowband, the relay decodes $s$ and forwards the decoded $\hat{s}$ in the second time slot. The narrowband transmit signal at the relay in the second time slot can be expressed as 
	\begin{equation}
		x[2] = \hat{s}.
	\end{equation}
	Thus, the narrowband received signals at the destination with RIS passive beamforming can be expressed as  
	\begin{equation}
		r^D[2] = \widetilde{g}^{\text{RD}}[2] \hat{s} + (\widetilde{\textbf{h}}^{\text{ID}}[2])'{\bf{\Phi}}[2]\widetilde{\textbf{h}}^{\text{RI}}[2]\hat{s} + n^D, \label{A3}
	\end{equation}
	where the AWGN at the destination is denoted by $n^D$, and the RIS passive beamforming matrix in the second time slot is denoted by ${\bf{\Phi}}[2] = \text{diag}\{\phi_1[2], \phi_2[2], \cdots, \phi_M[2]\}$ with $\phi_m[2] = e^{j\theta_m[2]},\theta_m[2]\in[0,2\pi],\forall m = 1,2,\cdots,M$.
	
	In the wideband OFDM, the frequency representation is the DFT of the time-domain signal. Thus, the frequency representation of the received signal at the relay in the time slot $1$ can be expressed as
	\begin{eqnarray}
		&&\!\!\!\!\!\!	\textbf{y}^R[1] = \mathcal{F}\{\widetilde{\textbf{g}}^\text{SR}[1]*\textbf{s} +  (\widetilde{\textbf{H}}^\text{IR}[1])'*{\bf{\Phi}}[1]\widetilde{\textbf{H}}^\text{SI}[1]*\textbf{s} + \widetilde{\textbf{n}}^R\} \nonumber \\
		&&\!\!\!\!\!\!	= \textbf{g}^\text{SR}[1]\textbf{s} + (\textbf{H}^\text{IR}[1])'{\bf{\Phi}}[1]\textbf{H}^\text{SI}[1]\textbf{s} + \textbf{n}^R,  \label{RSS1}
	\end{eqnarray}
	where for time-domain, $\widetilde{\textbf{g}}^\text{SR}[1] \triangleq [\widetilde{g}^\text{SR}_1[1];\cdots;\widetilde{g}^\text{SR}_L[1];\textbf{0}]$ with $L$ taps, $\widetilde{\textbf{H}}^\text{IR}[1] \triangleq [\widetilde{\textbf{h}}_1^\text{IR}[1],\cdots,\widetilde{\textbf{h}}_L^\text{IR}[1],\textbf{0}]$ with $L$ taps, and $\widetilde{\textbf{H}}^\text{SI}[1] \triangleq [\widetilde{\textbf{h}}_1^\text{SI}[1],\cdots,\widetilde{\textbf{h}}_L^\text{SI}[1],\textbf{0}]$ with $L$ taps; for frequency-domain, $\textbf{g}^\text{SR}[1] \triangleq [g^\text{SR}_1[1];\cdots;g^\text{SR}_N[1]]$ with $N$ subcarriers, $\textbf{H}^\text{IR}[1] \triangleq [\textbf{h}_1^\text{IR}[1],\cdots,\textbf{h}_N^\text{IR}[1]]$ with $N$ subcarriers, $\textbf{H}^\text{SI}[1] \triangleq [\textbf{h}_1^\text{SI}[1],\cdots,\textbf{h}_N^\text{SI}[1]]$ with $N$ subcarriers, and AWGN $\textbf{n}^R \sim \mathcal{CN}(\textbf{0},\sigma^2\textbf{I}_N)$. The frequency representation of the received signal at the destination in the time slot $2$ can be expressed as
	\begin{eqnarray}
		&&\!\!\!\!\!\!	\textbf{y}^D[2] = \mathcal{F}\{\widetilde{\textbf{g}}^\text{RD}[2]*\hat{\textbf{s}} +  (\widetilde{\textbf{H}}^\text{ID}[2])'*{\bf{\Phi}}[2]\widetilde{\textbf{H}}^\text{RI}[2]*\hat{\textbf{s}} + \widetilde{\textbf{n}}^D\} \nonumber \\
		&&\!\!\!\!\!\!	= \textbf{g}^\text{RD}[2]\hat{\textbf{s}} + (\textbf{H}^\text{ID}[2])'{\bf{\Phi}}[2]\textbf{H}^\text{RI}[2]\hat{\textbf{s}} + \textbf{n}^D,  \label{RSS2}
	\end{eqnarray}
	where for time-domain, $\widetilde{\textbf{g}}^\text{RD}[2] \triangleq [\widetilde{g}^\text{RD}_1[2];\cdots;\widetilde{g}^\text{RD}_L[2];\textbf{0}]$ with $L$ taps, $\widetilde{\textbf{H}}^\text{ID}[2] \triangleq [\widetilde{\textbf{h}}_1^\text{ID}[2],\cdots,\widetilde{\textbf{h}}_L^\text{ID}[2],\textbf{0}]$ with $L$ taps, and $\widetilde{\textbf{H}}^\text{RI}[2] \triangleq [\widetilde{\textbf{h}}_1^\text{RI}[2],\cdots,\widetilde{\textbf{h}}_L^\text{RI}[2],\textbf{0}]$ with $L$ taps; for frequency-domain, $\textbf{g}^\text{RD}[2] \triangleq [g^\text{RD}_1[2];\cdots;g^\text{RD}_N[2]]$ with $N$ subcarriers, $\textbf{H}^\text{ID}[2] \triangleq [\textbf{h}_1^\text{ID}[2],\cdots,\textbf{h}_N^\text{ID}[2]]$ with $N$ subcarriers, $\textbf{H}^\text{RI}[2] \triangleq [\textbf{h}_1^\text{RI}[2],\cdots,\textbf{h}_N^\text{RI}[2]]$ with $N$ subcarriers, and AWGN $\textbf{n}^D \sim \mathcal{CN}(\textbf{0},\sigma^2\textbf{I}_N)$.
	
	It is worth mentioning that 1) RIS generally can impose different passive beamforming designs over the two-time slots, thus having more degrees-of-freedom for improving the communication performance; 2) the propagation delay difference in the source-relay and source-RIS-relay links, as well as the relay-destination and relay-RIS-destination links, is marginal and can be seen as the same time.

	\subsection{Case-II: With RIS-destination link in Time Slot One}

	As shown in Fig. \ref{F22},  in Case-II, we propose a new design framework, where the RIS-destination link is considered in the first time slot. This is motivated by the fact for the considered relaying system, the decodable rate in the destination usually is slightly smaller than that in the relay, due to the user's random location and hence degrades link performance. To address this issue, the additional RIS-destination link in the first slot can be leveraged to improve the achievable rate at the destination by efficiently designing the RIS passive beamforming.  The narrowband received signal at the destination in the time slot $1$ can be expressed as  
	\begin{equation}
		r^D[1] =   (\widetilde{\textbf{h}}^{\text{ID}}[1])'{\bf{\Phi}}[1]\widetilde{\textbf{h}}^{\text{SI}}[1]  s + n_D. \label{B1}
	\end{equation}
	
	In the wideband OFDM, the frequency representation is the DFT of the time-domain signal. Thus, the frequency representation of the received signal at the destination in the time slot $1$ can be expressed as
	\begin{eqnarray}
		&&\!\!\!\!\!\!	\textbf{y}^D[2] = \mathcal{F}\{  (\widetilde{\textbf{H}}^\text{ID}[1])'*{\bf{\Phi}}[1]\widetilde{\textbf{H}}^\text{RI}[1]*\textbf{s} + \widetilde{\textbf{n}}^D\} \nonumber \\
		&&\!\!\!\!\!\!	=  (\textbf{H}^\text{ID}[1])'{\bf{\Phi}}[1]\textbf{H}^\text{RI}[1]\textbf{s} + \textbf{n}^D,  \label{RSS3}
	\end{eqnarray}
	where for time-domain,  $\widetilde{\textbf{H}}^\text{ID}[1] \triangleq [\widetilde{\textbf{h}}_1^\text{ID}[1],\cdots,\widetilde{\textbf{h}}_L^\text{ID}[1],\textbf{0}]$ with $L$ taps, and $\widetilde{\textbf{H}}^\text{RI}[1] \triangleq [\widetilde{\textbf{h}}_1^\text{RI}[1],\cdots,\widetilde{\textbf{h}}_L^\text{RI}[1],\textbf{0}]$ with $L$ taps; for frequency-domain,  $\textbf{H}^\text{IR}[1] \triangleq [\textbf{h}_1^\text{IR}[1],\cdots,\textbf{h}_N^\text{IR}[1]]$ with $N$ subcarriers, $\textbf{H}^\text{SI}[1] \triangleq [\textbf{h}_1^\text{SI}[1],\cdots,\textbf{h}_N^\text{SI}[1]]$ with $N$ subcarriers, and AWGN $\textbf{n}^D \sim \mathcal{CN}(\textbf{0},\sigma^2\textbf{I}_N)$.
	In contrast to Case-I, this signal takes into account. As we will see,   the SNR expression at the destination differs from that in Case-I by adding (8).


	\section{Problem Formulation}
	
	We first obtain the achievable rate under the DF protocol for the two cases.  Specifically, based on \eqref{RSS1}, both for Case-I and II, SNR at the relay and subcarrier $p$ can be  written as
	\begin{equation}
		\text{SNR}^{R}_p = \frac{P_1|g^{\text{SR}}_p[1] + (\textbf{h}^{\text{IR}}_p[1])'{\bf{\Phi}}[1]\textbf{h}^{\text{SI}}_p[1]|^2}{\sigma^2}, \label{A4}
	\end{equation}
	where $P_1$ denotes the per subcarrier transmit power of the source, and $\sigma^2$ denotes the AWGN power on each subcarrier.
	Based on \eqref{RSS2}, for Case-I, SNR at the destination and subcarrier $q$  can be written as 
	\begin{equation}
		\text{SNR}^{D}_{q} = \frac{P_2|g^{\text{RD}}_q[2]  + (\textbf{h}^{\text{ID}}_q[2])'{\bf{\Phi}}[2]\textbf{h}^{\text{RI}}_q[2]|^2}{\sigma^2},
		\label{A5}
	\end{equation}
	where $P_2$ denotes the per subcarrier transmit power of the relay.
	Based on (6) and (8), for Case-II, by exploiting the stored received signals in time slots $1$ and $2$, we can express the SNR at the destination and subcarrier $p$ and $q$ as 
	\begin{eqnarray}
		&& \text{SNR}^{D}_{p,q} = \frac{P_1|(\textbf{h}^{\text{ID}}_p[1])'{\bf{\Phi}}[1]\textbf{h}^{\text{SI}}_p[1]|^2}{\sigma^2} \nonumber \\ &&+ \frac{P_2|g^{\text{RD}}_q[2]  
			+ (\textbf{h}^{\text{ID}}_q[2])'{\bf{\Phi}}[2]\textbf{h}^{\text{RI}}_q[2]|^2}{\sigma^2}. \label{B3}
	\end{eqnarray} Note that there is no subcarrier transfer in RIS. 
	For simplicity, we define an indicator $k\in\{0,1\}$ for the two cases. As such, we can express the SNR at the destination and subcarrier $p$ and $q$ for Case-I and II as  
	\begin{eqnarray}
		&& \text{SNR}^{D}_{p,q}(k) = k\frac{P_1|(\textbf{h}^{\text{ID}}_p[1])'{\bf{\Phi}}[1]\textbf{h}^{\text{SI}}_p[1]|^2}{\sigma^2} \nonumber \\ &&+ \frac{P_2|g^{\text{RD}}_q[2]  
			+ (\textbf{h}^{\text{ID}}_q[2])'{\bf{\Phi}}[2]\textbf{h}^{\text{RI}}_q[2]|^2}{\sigma^2}, \label{UB3}
	\end{eqnarray}
	where $k=0$ indicates Case-I and $k=1$ stands for Case-II.
	
	Our goal is to maximize the sum achievable rate at the destination overall subcarrier pairs by jointly optimizing the RIS passive beamforming and subcarrier matching.
	Mathematically, this joint optimization problem for Case-I ($k=0$) and II ($k=1$)  can be formulated as follows:
	\begin{subequations}
		\begin{eqnarray}
			\mathcal{P}_1 \quad \max_{\{{\bm{\phi}}[t],  
				x_{p,q}\}} && \frac{\Delta}{2T} \sum_{p=1}^N \sum_{q=1}^N \log_2 \left(1 + \min\{\text{SNR}^{R}_p,  \right. \nonumber \\ 
			&& \left. x_{p,q}\text{SNR}^{D}_{p,q}(k)\} \right) \nonumber \\ 
			\text{s.t.} && \sum_{p=1}^N x_{p,q} \le 1,\quad \forall q, \, \\
			&& \sum_{q=1}^N x_{p,q} \le 1,\quad \forall p, \\
			&& |\phi_m[t]| = 1, \quad \forall m, t, \label{PPC} \\
			&& x_{p,q} \in \{0,1\}, \quad \forall p,q,
		\end{eqnarray}
	\end{subequations}
	where $\text{SNR}^{R}_p$ and $\text{SNR}^{D}_{p,q}$ are given in \eqref{A4} and \eqref{UB3}, respectively, and $x_{p,q} \in \{0,1\}$ is the binary subcarrier matching variable, indicating whether or not the symbol modulated on subcarrier $p$ is matched to subcarrier $q$ for relay transmission. Due to the orthogonality of the OFDM signal, the subcarrier matching should be exclusive. That is,  subcarrier $p$ of source transmission can  be matched to at most one subcarrier for relay transmission (i.e., $\sum_{q=1}^N x_{p,q} \le 1$), and subcarrier $q$ for relay transmission can be only assigned to at most one subcarrier for source transmission (i.e., $\sum_{p=1}^N x_{p,q} \le 1$). The optimization Problem $\mathcal{P}_1$ is an MINLP problem, which is difficult to deal with, due to the non-convexity of binary variables. Additionally, the uni-modulus phase constraint \eqref{PPC} renders the problem more challenging to solve.
	
	\section{Branch-and-Bound-based Alternating Optimization Algorithm}

\begin{figure}[t]
		\centering
		\includegraphics[width=2.8in]{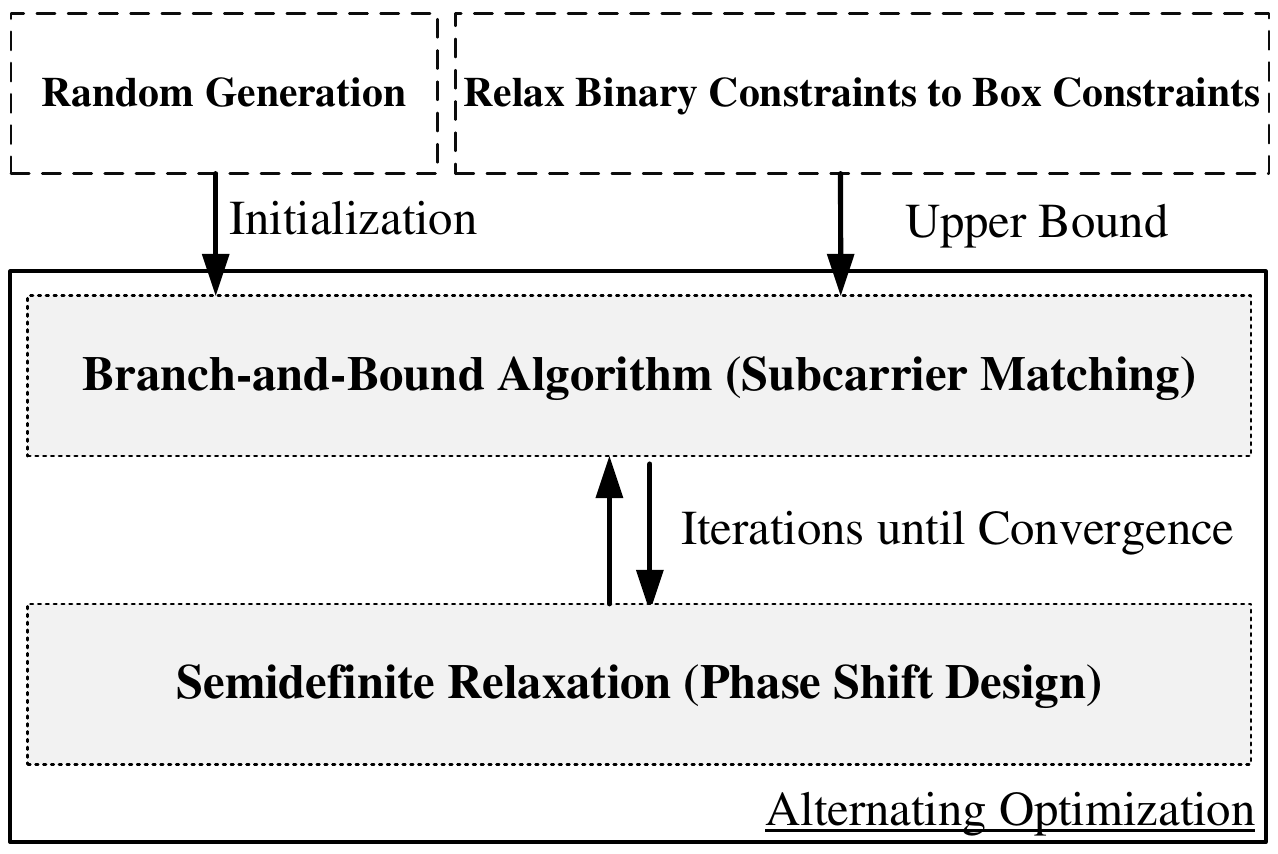}
		\caption{Framework of the BnB-based alternating optimization algorithm.} \label{frame}
	\end{figure}

	To tackle the issue in the coupling of subcarrier matching and RIS passive beamforming,  we first divide the MINIP Problem $\mathcal{P}_1$ into a subcarrier matching subproblem and a RIS passive beamforming subproblem, and then alternatively solve these two subproblems until convergence. Specifically, the subcarrier matching problem is a binary integer problem. A naive way to solve the subcarrier matching problem is using an exhaustive search.  However, the search space of subcarrier matching is exponentially large (i.e, ${\cal{O}}(2^{N^2})$), which is computationally prohibitive. We, therefore, devise a BnB algorithm to optimally solve the subcarrier matching problem with reduced complexity. Next, the RIS passive beamforming problem is addressed by SDR. For ease of explanation, we illustrate the framework of the BnB-based alternating optimization algorithm in Fig. \ref{frame}. The details of the proposed algorithms are elaborated below.

	\subsection{Branch-and-Bound-based Subcarrier Matching}

	To begin with, we define a living pool ${\cal{V}}$ as the solution set of $\{x_{p,q}\}$, that has not been explored, and the incumbent $I$ as the current best objective value that has been obtained. We initialize the incumbent $I$ by finding a feasible solution that satisfies MINIP Problem $\mathcal{P}_1$ given RIS passive beamforming. Next, we exhaust the living pool ${\cal{V}}$ by popping up its elements one by one. The popup element is branched into two subproblems by letting $x' = 0$ and $x' = 1$, where $x'$ denotes the branched element. These two subproblems are relaxed to upper bound problems, which will be elaborated on later. If the objective value of the upper bound problem is not larger than the incumbent $I$ or the upper bound problem is infeasible, then the corresponding branching (i.e., $x' = 0$ or $x' = 1$) will be discarded. If the optimal objective value of the upper bound problem is larger than the incumbent $I$ and optimal solutions of the upper bound problem are also feasible to the MINIP problems given RIS passive beamforming, the incumbent $I$ will be replaced by this objective value. If the optimal objective value of the upper bound problem is larger than the incumbent $I$, while optimal solutions of the upper bound problem are infeasible to the MINIP problems given RIS passive beamforming, this branch will be saved and subsequent branching will proceed. The BnB algorithm ends until the living pool ${\cal{V}}$ is empty. More specifically, the brand-and-bound algorithm includes the following three procedures:
	\begin{figure}[t]
		\centering
		\includegraphics[width=3.55in]{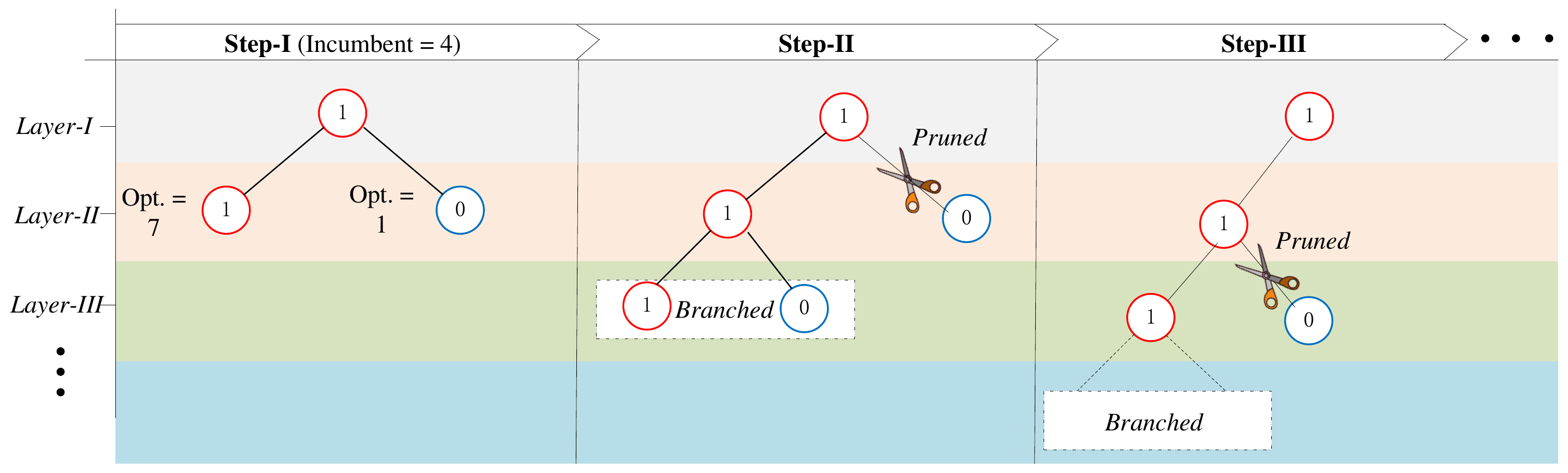}
		\caption{Example of branching, bounding, and pruning, where the branch is pruned since its optimal objective values are less than the incumbent.} \label{Branch}
	\end{figure}
	\begin{enumerate}
		\item \textbf{Branching}: The BnB algorithm solves MINIP Problem $\mathcal{P}_1$ given RIS passive beamforming by recursively dividing it into subproblems with fewer binary variables. The first two subproblems are made by selecting a binary variable from the living pool ${\cal{V}}$ and fixing it to zero in the first subproblem and to one in the	second. Each of the two subproblems may be further divided into two more subproblems by fixing a second binary variable. This branching process produces a binary tree of subproblems. 
		
		\item \textbf{Bounding}: This phase aims to evaluate whether or not the branch should be pruned.  The upper bound problem should be carefully designed so that it can be globally optimally solved, which serves as an upper bound.
		
		\item \textbf{Pruning}: This procedure aims to accelerate the branching process, which prunes the impossible branches. Denote the optimal objective value of subproblems 1 and 2 as $S_1$ and $S_2$, respectively. Additionally, if the subproblems are infeasible, the optimal objective value will be set to $-{\cal{1}}$. The pruning rule is given as follows: 1) If $S_i \le I,\,i=1,2$, the branch of subproblem $i$ will be pruned. 2) If $S_i > I, \,i=1,2$ and the optimal solutions of subproblem $i$ are feasible to MINIP Problem $\mathcal{P}_1$ given RIS passive beamforming, the incumbent $I$ will be reset (i.e., $I = \max\{S_1,S_2\}$) and the corresponding branch  will keep branching.  3) If $S_i > I, \,i=1,2$ and the optimal solutions of subproblem $i$ are infeasible to MINIP Problem $\mathcal{P}_1$ given RIS passive beamforming, the corresponding branch will keep branching.  	
	\end{enumerate}
	
	A toy example of branching, bounding, and pruning is illustrated in Fig. \ref{Branch}. As for the bounding procedure, we need to obtain the upper bound problem of MINIP Problem $\mathcal{P}_1$ given RIS passive beamforming.
	In the following, based on the box relaxation, the upper bound problem of MINIP Problem $\mathcal{P}_1$ given RIS passive beamforming can be designed as follows:
	
	\begin{algorithm}[t]
		\caption{The BnB Algorithm for Solving Problem $\mathcal{P}_1$ Given RIS Passive Beamforming}
		\begin{algorithmic}[1]
			\State \textbf{Initialize}: Random initialization and set living pool 
			$\mathcal{V}$  $=\{x_{1,1},x_{1,2},\cdots,x_{1,N},x_{2,1},x_{2,2},\cdots,x_{2,N},\cdots,x_{N,N}\}$.
			\State \textbf{Repeat}:		
			\State \quad  Pick the rightest element, denoted by $\hat{x}$, from the \State \quad living pool $\mathcal{V}$.
			\State \quad \textbf{Branch}  $\hat{x}$
			\State \quad \quad Let $\hat{x} = 0$. Solve Problem $\mathcal{P}_3$ by CVX,  where
			\State \quad \quad  optimal objective value  are denoted  by $v_\text{opt}^{\mathcal{P}_3}$. 
			
			\State \quad  \quad \textbf{If} $v_\text{opt}^{\mathcal{P}_3} \le I$ 
			\State \quad  \quad  \quad Prune $\hat{x} = 0$		 
			\State \quad  \quad  \textbf{Otherwise}
			\State \quad  \quad  \quad \textbf{If} the optimal solutions of Problem $\mathcal{P}_3$
			\State \quad  \quad  \quad is feasible  to Problem $\mathcal{P}_1$.  
			\State \quad  \quad  \quad \quad $I = v_\text{opt}^{\mathcal{P}_3}$ 
			\State \quad  \quad  \quad \textbf{End If}
			\State \quad  \quad  \quad Fix $\hat{x} = 0$
			\State  \quad  \quad \textbf{End If}
			\State  \quad  \quad  Let $\hat{x} = 1$. Solve Problem $\mathcal{P}_3$ by CVX.
			\State \quad  \quad  \textbf{If} $v_\text{opt}^{\mathcal{P}_3} \le I$ 
			\State \quad  \quad  \quad Prune $\hat{x} = 1$
			\State \quad  \quad \textbf{Otherwise}
			\State \quad  \quad  \quad \textbf{If} the optimal solutions of Problem $\mathcal{P}_3$ 
			\State \quad  \quad  \quad  is feasible  to Problem $\mathcal{P}_1$.  
			\State \quad  \quad  \quad \quad $I = v_\text{opt}^{\mathcal{P}_3}$ 
			\State  \quad  \quad  \quad \textbf{End If}
			\State \quad  \quad  \quad Fix $\hat{x} = 1$
			\State \quad  \quad  \textbf{End If}
			\State \quad  $\mathcal{V} = \mathcal{V}\backslash\{\hat{x}\}$
			\State \textbf{Until}: $\mathcal{V}$ is empty
		\end{algorithmic}
	\end{algorithm}
	
	Firstly, we equivalently transform Problem $\mathcal{P}_1$ as follows by introducing slack variables $\{\alpha_{p,q}\}$, which represents the minimal SNR:
	\begin{subequations}
		\begin{eqnarray}
			\mathcal{P}_2 \max_{\{x_{p,q},\alpha_{p,q}\}} && \!\!\!\!\!\!\! \frac{\Delta}{2T} \sum_{p=1}^N \sum_{q=1}^N \log_2 \left(1 + \alpha_{p,q} \right) \nonumber \\
			\text{s.t.} && \!\!\!\!\!\!\! \alpha_{p,q} \le  \text{SNR}^{R}_p, \quad \forall p,q, \\
			&& \!\!\!\!\!\!\! \alpha_{p,q} \le x_{p,q}\text{SNR}^{D}_{p,q}(k), \quad \forall p,q, \\
			&& \!\!\!\!\!\!\! \sum_{p=1}^N x_{p,q} \le 1,\quad \forall q, \\
			&& \!\!\!\!\!\!\! \sum_{q=1}^N x_{p,q} \le 1,\quad \forall p, \\
			&& \!\!\!\!\!\!\! x_{p,q} \in \{0,1\}, \quad \forall p,q,
		\end{eqnarray}
	\end{subequations}
	where $\text{SNR}_p^R$ and $\text{SNR}_{p,q}^D(k)$ are given in \eqref{A4} and \eqref{UB3}, respectively. Note that given RIS passive beamforming, $\text{SNR}_p^R$ and $\text{SNR}_{p,q}^D(k)$ are fixed. Based on the box relaxation, we relax $x_{p,q} \in \{0,1\}$ to $0 \le x_{p,q} \le 1$. As such, the upper bound problem of Problem $\mathcal{P}_2$ is given by
	\begin{subequations}
		\begin{eqnarray}
			\mathcal{P}_3	\max_{\{x_{p,q},\alpha_{p,q}\}} && \!\!\!\!\!\!\! \frac{\Delta}{2T} \sum_{p=1}^N \sum_{q=1}^N \log_2 \left(1 + \alpha_{p,q} \right) \nonumber \\
			\text{s.t.} && \!\!\!\!\!\!\!  \text{(14a)}-\text{(14d)}, \nonumber \\
			&& \!\!\!\!\!\!\! 0 \le x_{p,q} \le 1, \quad \forall p,q.
		\end{eqnarray}
	\end{subequations}
	Since the binary variables have been relaxed to continuous variables, this problem is convex and can be solved by CVX.

	Utilizing the above three techniques (i.e., branching, bounding, pruning), we summarize the proposed BnB algorithm in Algorithm 1, which optimally solves the MINIP Problem $\mathcal{P}_1$ given RIS passive beamforming \cite{405}. Next, we design the RIS passive beamforming given the subcarrier matching.

	\subsection{SDR-based RIS Passive Beamforming}

	Given the subcarrier matching, RIS passive beamforming can be obtained by the following SDR procedure. First of all, we introduce slack variables $\{\alpha_{p,q}\}$. As such, given the subcarrier matching, Problem $\mathcal{P}_1$ can be equivalently transformed into 
	\begin{subequations}
		\begin{eqnarray}
			\mathcal{P}_4 \max_{\{\textbf{v}_t, \alpha_{p,q}\}} && \!\!\!\!\!\!\!\!\!\! \frac{\Delta}{2T} \sum_{p=1}^N \sum_{q=1}^N \log_2 \left(1 + \alpha_{p,q} \right) \nonumber \\
			\text{s.t.} && \!\!\!\!\!\!\!\!\!\! \alpha_{p,q} \le \frac{P_1|g^{\text{SR}}_p[1] + \textbf{v}'_1\textbf{a}_{p}|^2}{\sigma^2},\,\,\, \forall p,q, \label{VVVV1}\\
			&& \!\!\!\!\!\!\!\!\!\! \alpha_{p,q} \le x_{p,q}\left(k\frac{P_1|\textbf{v}_1'\textbf{c}_{p,q}|^2}{\sigma^2} \right. \nonumber \\
			&& \!\!\!\!\!\!\!\!\!\! \left. + \frac{P_2|g^{\text{RD}}_q[2]  + \textbf{v}'_2\textbf{b}_q|^2}{\sigma^2}\right), \,\,\, \forall p,q, \label{VVVV3}\\
			&& \!\!\!\!\!\!\!\!\!\! |\phi_m[t]| = 1, \quad \forall m,t, \label{VVVV4}  
		\end{eqnarray}
	\end{subequations} 
	Towards formulating Problem $\mathcal{P}_4$ as a semidefinite programming (SDP) problem, we introduce $\widetilde{\textbf{v}}_1 = [\textbf{v}_1;t_1]$ with a slack variable $t_1$, since $|g^{\text{SR}}_p[1] + \textbf{v}'_1\textbf{a}_{p}|^2 = |g^{\text{SR}}_p[1]|^2 + \textbf{v}'_1\textbf{a}_p(g^{\text{SR}}_p[1])^* + g^{\text{SR}}_p[1]\textbf{a}_p'\textbf{v}_1 + \textbf{v}'_1 \textbf{a}_p \textbf{a}_p' \textbf{v}_1$ and $|\textbf{v}_1'\textbf{c}_{p,q}|^2 = \textbf{v}'_1 \textbf{c}_{p,q} \textbf{c}_{p,q}' \textbf{v}_1$, we can re-write the constraint \eqref{VVVV1} as 
	\begin{equation}
		\alpha_{p,q} \le \frac{P_1}{\sigma^2}\left( \widetilde{\textbf{v}}'_1\underbrace{\begin{bmatrix}
				\textbf{a}_p\textbf{a}_p' & \textbf{a}_p(g^{\text{SR}}_p[1])^* \\
				g^{\text{SR}}_p[1]\textbf{a}_p' & 0 	
		\end{bmatrix}}_{\text{denoted by}\,\,\textbf{A}_p}\widetilde{\textbf{v}}_1 + |g^{\text{SR}}_p[1]|^2\right). \label{TT1}
	\end{equation} 
	By introducing $\widetilde{\textbf{v}}_2 = [\textbf{v}_2;t_2]$ with a slack variable $t_2$, due to $|g^{\text{RD}}_q[2]  + \textbf{v}'_2\textbf{b}_q|^2 = |g_q^\text{RD}[2]|^2 + \textbf{v}'_2\textbf{b}_q(g_p^\text{RD}[2])^* + g_p^\text{RD}[2]\textbf{b}_q'\textbf{v}_2+ \textbf{v}'_2\textbf{b}_q\textbf{b}'_{q}\textbf{v}_2$, we re-write  the constraint \eqref{VVVV3} as  \eqref{ETE}.
	\begin{figure*}
		\begin{eqnarray}
			&&\alpha_{p,q} \le x_{p,q}\left(kP_1\widetilde{\textbf{v}}_1' \underbrace{\begin{bmatrix} 
					\textbf{c}_{p,q}\textbf{c}_{p,q}' & \textbf{0} \\
					\textbf{0} & 0
			\end{bmatrix}}_{\text{denoted by}\,\,\textbf{C}_{p,q}} \widetilde{\textbf{v}}_1/\sigma^2 + P_2\left(\widetilde{\textbf{v}}_2' \underbrace{\begin{bmatrix}
					\textbf{b}_q\textbf{b}_q' & \textbf{b}_q(g_q^\text{RD}[2])^* \\
					g_q^\text{RD}[2] \textbf{b}_q' & 0
			\end{bmatrix}}_{\text{denoted by}\,\,\textbf{B}_{q}} \widetilde{\textbf{v}}_2    + |g_p^\text{RD}[2]|^2\right)/\sigma^2\right). \label{ETE}
		\end{eqnarray} 
		\hrule
	\end{figure*}
	
	In order to transform the problem into a SDP, we introduce semidefinite matrices ${\bm{\Phi}}_t,\,t=1,2$. Considering the non-convex constraint that the rank of $\{{\bm{\Phi}}_t\}$ should be one, we have $\text{Tr}\{{\bm{\Phi}}_1 \textbf{A}_{p}\} = {\bm{\Phi}}_1'\textbf{A}_{p}{\bm{\Phi}}_1$, $\text{Tr}\{{\bm{\Phi}}_2 \textbf{B}_{q}\} = {\bm{\Phi}}_2'\textbf{B}_{q}{\bm{\Phi}}_2$, and $\text{Tr}\{{\bm{\Phi}}_1 \textbf{C}_{p,q}\} = {\bm{\Phi}}_1'\textbf{C}_{p}{\bm{\Phi}}_1$. Dropping the rank-1 constraints, we finally arrive at the following SDP problem:
	\begin{subequations}
		\begin{eqnarray}
			\mathcal{P}_5 \max_{\{{\bm{\Phi}}_t, \alpha_{p,q}\}} && \!\!\!\!\!\!\!\!\!\!\!\! \frac{\Delta}{2T} \sum_{p=1}^N \sum_{q=1}^N \log_2 \left(1 + \alpha_{p,q} \right) \nonumber \\
			\text{s.t.} && \!\!\!\!\!\!\!\!\!\!\!\! \alpha_{p,q} \le  P_1\frac{ \text{Tr}\{{\bm{\Phi}}_1 \textbf{A}_p\} + |{\textbf{g}^{\text{SR}}}'\textbf{f}_p|^2}{\sigma^2}, \forall p,q,\\
			&&\!\!\!\!\!\!\!\!\!\!\!\! \alpha_{p,q} \le x_{p,q}\left(kP_1\frac{\text{Tr}\{{\bm{\Phi}}_1 \textbf{C}_{p}\}}{\sigma^2} \right.\nonumber \\
			&& \!\!\!\!\!\!\!\!\!\!\!\!
			\left. + P_2\frac{\text{Tr}\{{\bm{\Phi}}_2 \textbf{B}_{q}\} + |{\textbf{h}^\text{RD}}'\textbf{f}_q|^2}{\sigma^2}\right),  \,\,  \forall p,q, \\
			&& \!\!\!\!\!\!\!\!\!\!\!\! {\bm{\Phi}}_t(m,m) = 1, \quad \forall m,t, \\
			&& \!\!\!\!\!\!\!\!\!\!\!\! {\bm{\Phi}}_t  \succeq \textbf{0}, \quad \forall t.
		\end{eqnarray}
	\end{subequations}
	The above SDP problem can be optimally solved by CVX. After obtaining the optimal solutions, the RIS passive beamforming can be derived from Gaussian randomization, given in Appendix D.  Based on \cite{406}, if the Gaussian randomization has a sufficiently large number of randomizations, it guarantees an $\pi/4$ -approximation of the optimal objective value.
	
	\subsection{The Overall Algorithm and Complexity Analysis}
	
	Based on the aforementioned subcarrier matching and RIS passive beamforming approaches, we can alternatively optimize subcarrier matching and RIS passive beamforming until convergence. The proposed BnB-based alternation optimization algorithm is given in Algorithm 2. Moreover, the convergence of Algorithm 2 is analyzed below.   
	
	\textbf{Proposition 1}: Algorithm 2  is guaranteed to converge

	\begin{IEEEproof}
		Please refer to Appendix \ref{AppendixB}.
	\end{IEEEproof}
	
	The worst-case computational complexity of Algorithm 2 is summarized below.
	\begin{algorithm}[t] 
		\caption{The BnB-based Alternation Optimization Algorithm for Solving MINIP Problem $\mathcal{P}_1$}
		\begin{algorithmic}[1]
			\State Randomly initialize the reflecting element matrices, and $\text{iter}=1$.
			\State \textbf{Repeat}:	 
			\State \quad Run Algorithm 1 \% \textit{Subcarrier Matching}
			\State \quad Solve Problem $\mathcal{P}_5$ by CVX \% \textit{Passive Beamforming}  
			\State \quad Update $\text{iter}=\text{iter}+1$ 
			\State \textbf{Until}: The increase of objective is below a threshold.
		\end{algorithmic}
	\end{algorithm}
	
	\textbf{Proposition 2}:  The worst-case computational complexity of Algorithm 2 is $\mathcal{O}(F(M^{3.5} + X_1(N^2)^{3.5} + X_22^{N^2}))$, where $F$ is number of rounds needed for convergence, $X_1$ is number of times computing the upper bound problem, $X_2$ is a parameter related to pruning.
	\begin{IEEEproof}
		Please refer to Appendix \ref{AppendixC}.
	\end{IEEEproof}

	\section{Difference-of-Convex Penalty-based Algorithm}
	
	Due to the exponential complexity of the BnB algorithm, the proposed BnB-based alternating optimization algorithm may be hard to implement especially when the number of subcarriers is large \cite{405}. To tackle this challenge, we resort to the difference-of-convex penalty-based approach, which has a much lower complexity. This approach equivalently transforms the binary variables into continuous variables by introducing a concave penalty \cite{402}. Next, we first propose the difference-of-convex penalty-based problem formulation and algorithm and analyze the convergence and complexity of the algorithm.

	\subsection{Difference-of-Convex Penalty-based Formulation}
	
	According to difference-of-convex penalty-based method \cite{402}, the binary variables $x_{p,q} \in \{0,1\}$, can be relaxed to continuous variables $0 \le x_{p,q} \le 1$, without loss of optimality, if the following penalty function with a suitable $\eta > 0$ is added into the original objective function:
	\begin{equation}
		\psi(\textbf{x}) = \eta\sum_{p=1}^N\sum_{q=1}^N x_{p,q}(1 - x_{p,q}). \label{penalty}	
	\end{equation}
	According to \cite[Proposition 1]{402}, with the penalty term \eqref{penalty}, there exists a  
	$\widetilde{\eta} > 0$ such that, for any  $\eta \in [0,\widetilde{\eta}]$, the original problem and penalty-based problem have the same optimal solutions. It can be seen that the penalty will be small if $x_{p,q}$ is close to $1$ or $0$. Also, the weight of penalty $\eta$ should be chosen in the same order as the original objective so that the relaxed continuous problem can approximate the original binary problem. The challenge of applying the penalty to the Problem $\mathcal{P}_1$ arises from the concavity of $-x_{p,q}^2$. Therefore, we propose to leverage the difference-of-convex procedure to sequentially convexify the concave function by a convex surrogate function \cite{407}. The proposed convex surrogate function is given by the following proposition:
	
	\textbf{Proposition 3}: The function of $-x^2$ can be upper bounded as follow:
	\begin{equation}
		-x^2 \le x' - 2 x' x,
	\end{equation}
	where $x'$ denotes the optimal $x$ obtained from the last iteration. 
	
	\begin{IEEEproof}
		The concave function is upper bounded by the first-order Taylor's expansion of $-x^2$, which is written as
		\begin{equation}
			-x^2 \le -2x'(x-x') - x'^2 = x'^2 - 2x'x,
		\end{equation}
		where Taylor's expansion is expanded at $x'$. 
	\end{IEEEproof}
	
	Applying the penalty term \eqref{penalty}, we present the relaxed continuous problem of Problem $\mathcal{P}_1$ given RIS passive beamforming below.
	\begin{subequations}
		\begin{eqnarray}
			\max_{\{x_{p,q},\alpha_{p,q}\}} && \frac{\Delta}{2T}\sum_{p=1}^N \sum_{q=1}^N \log_2 (1+\alpha_{p,q})  \nonumber \\
			&&  - \eta\sum_{p=1}^N\sum_{q=1}^N x_{p,q}(1 - x_{p,q})\nonumber \\
			\text{s.t.}
			&& \alpha_{p,q} \le A_p, \quad \forall p,q,  \\
			&& \alpha_{p,q} \le x_{p,q} (kC_p + B_{q}), \quad \forall p,q, \\ 
			&& \sum_{p=1}^N x_{p,q} \le 1, \quad \forall q,   \\
			&& \sum_{q=1}^N x_{p,q} \le 1, \quad \forall p,   \\
			&&  0 \le x_{p,q} \le 1, \quad \forall p,q,  
		\end{eqnarray}
	\end{subequations} 
	where \begin{subequations}
		\begin{eqnarray}
			&&	A_p = \frac{P_1|g^{\text{SR}}_p[1] + {\bm{\Phi}}'_1\textbf{A}_{p}|^2}{\sigma^2}, \\
			&&	B_q = \frac{P_2|g^{\text{RD}}_q[2]  + {\bm{\Phi}}'_2\textbf{B}_q|^2}{\sigma^2},\\
			&&		C_p = \frac{P_1|{\bm{\Phi}}_1'\textbf{C}_{p}|^2}{\sigma^2}.
		\end{eqnarray}
	\end{subequations} 
	Using Proposition 3, we can obtain the $i^\text{th}$-iteration convex Problem in the difference-of-convex procedure, given by
	\begin{eqnarray}
		\mathcal{P}_6 \max_{\{x_{p,q},\alpha_{p,q}\}} && \!\!\!\!\!\!\!\!\!\frac{\Delta}{2T}\sum_{p=1}^N \sum_{q=1}^N \log_2 (1+\alpha_{p,q}) \nonumber \\ &&\!\!\!\!\!\!\!\!\! - \eta\sum_{p=1}^N\sum_{q=1}^N ( x_{p,q}+ (x^{i-1}_{p,q})^2 
		- 2x_{p,q}^{i-1}x_{p,q}) \nonumber \\
		\text{s.t.} &&\!\!\!\!\!\!\!\!\!  \text{(20a)}-\text{(20e)}.  \nonumber 
	\end{eqnarray}

	\begin{figure*}[t]	 
		\centering
		\includegraphics[width=5in]{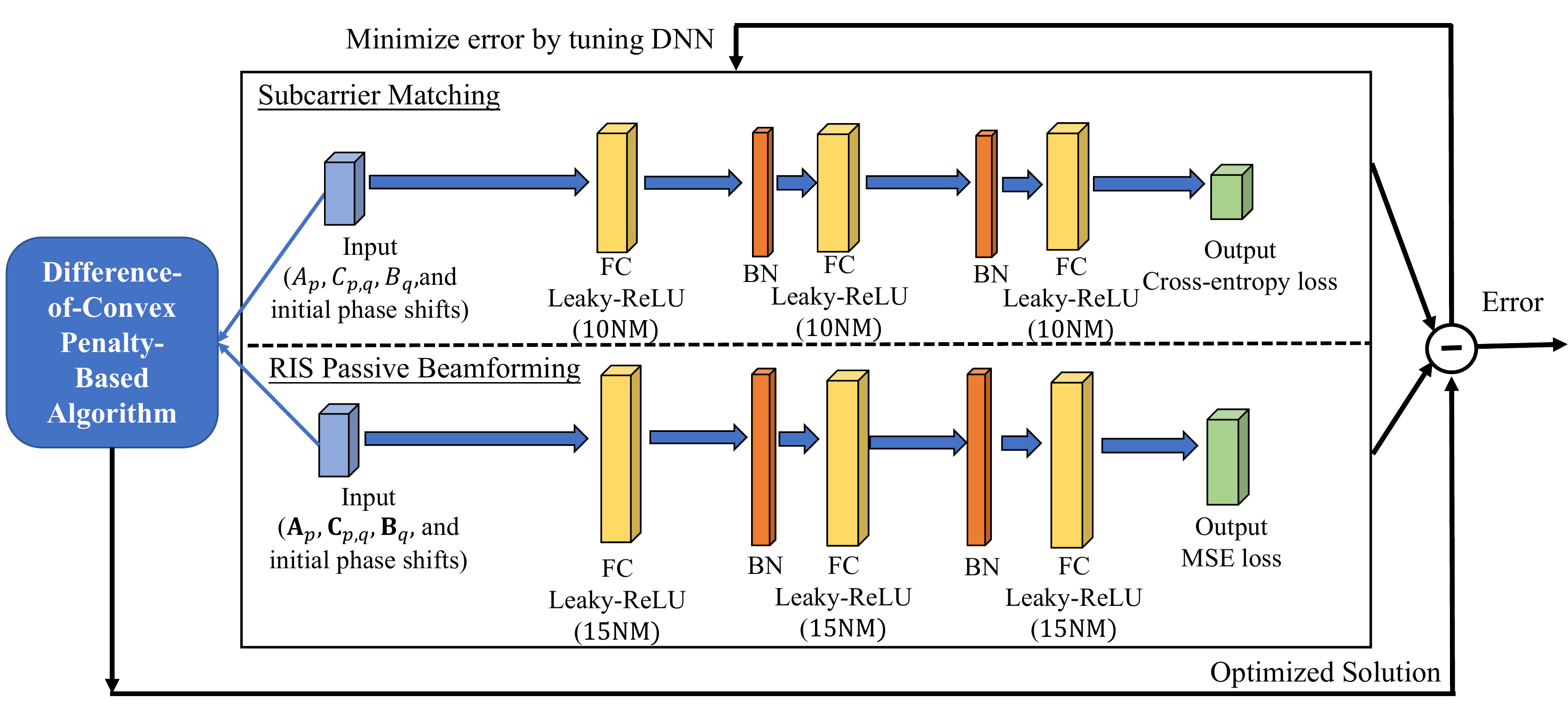}
		\caption{Illustration of the proposed DNN structures and DNN training process.} \label{ELF}
	\end{figure*}
	
	\begin{algorithm}[t] 
		\caption{The Difference-of-Convex Penalty-based Algorithm for Solving MINIP Problem $\mathcal{P}_1$}
		\begin{algorithmic}[1]
			\State Randomly initialize  $\{{\bm{\phi}}^\text{1}[t]\}$, and set $\text{iter}=1$.
			\State \textbf{Repeat}:
			\State \quad Diagonal initialization: $x_{p,q} = 1$ if $p=q,$ and $x_{p,q} = 0$ 
			\State \quad if  $p\ne q$. 
			\State \quad Set $i=1$.
			\State \quad \textbf{Repeat}:
			\State \quad \quad Solve Problem $\mathcal{P}_6$ by CVX \% \textit{Subcarrier Matching}
			\State \quad \quad Update $i=i+1$
			\State \quad \textbf{Until}: The increase of objective is below a
			threshold. 
			\State \quad Solve Problem $\mathcal{P}_5$ by CVX \% \textit{Passive Beamforming}  
			\State \quad Update $\text{iter}=\text{iter}+1$ 
			\State \textbf{Until}: The increase of objective is below a threshold.
		\end{algorithmic}
	\end{algorithm}

	
	
	\subsection{The Overall Algorithm and Complexity Analysis}
	Finally, we summarize the proposed difference-of-convex penalty-based algorithm in Algorithm 3. The convergence and complexity of Algorithm 3 are analyzed below. Firstly, the convergence of inner iterations in Algorithm 3 is guaranteed in \cite{407} by the difference-of-convex procedure. Secondly, the convergence of outer iterations can be proven in a similar way to Proposition 1, which is omitted for simplicity.  Since the objective functions of MINIP Problem $\mathcal{P}_1$ is upper-bounded, Algorithm 3 is guaranteed to converge.
	The worst-case computational complexity of Algorithm 3 is given by the following proposition:
	
	\textbf{Proposition 4}: The worst-case computational complexity of Algorithm 3 is $\mathcal{O}(K_2(K_1N^{3.5}+M^{3.5}))$, where $K_1,K_2$ are number of inner and outer iterations, respectively.  
	
	\begin{IEEEproof}
		Please refer to Appendix \ref{AppendixD}.
	\end{IEEEproof}
	
	In practice, $K_1, K_2$ are usually not very large. According to Proposition 3, the worst-case computational complexity of Algorithm 3 is polynomial, rather than exponential. Hence,  Algorithm 3 has a much lower complexity than that of the BnB-based alternating optimization algorithm.
	
	\section{Learning-to-Optimize Approach}
	
	It is worth mentioning that when the wireless channel is fast fading, the algorithm running time should be extremely low, to catch up with the changing speed of the wireless channel. Therefore, to further reduce the complexity of optimization, the learning-to-optimize approach is adopted \cite{400}, which trains deep neural networks (DNNs) to learn from the inputs and outputs of optimization algorithms.  The deep-learning-based algorithm has a very low computation complexity in practice, which will be evaluated theoretically and by simulations.

	\subsection{Structure of Proposed DNNs}
	
	The problem for joint optimization of the subcarrier matching and RIS passive beamforming renders two types of variables, where the solutions of subcarrier matching are binary variables and the solutions of RIS passive beamforming are continuous and complex variables. Since binary variables play as the decision indicator while continuous variables do not, the loss functions for binary and continuous variables should be different. This motivates us to propose two DNNs\footnote{One may consider the multi-task learning (MTL) to jointly optimize subcarrier matching and RIS passive beamforming using one DNN. However, the advantage of MTL is based on the tasks’ relatedness measured by the degree of parameter and feature sharing, which is limited for these two tasks. Moreover, jointly learning two unrelated tasks using one model will potentially increase the noise and decrease the effectiveness \cite{li2017better}. This thus motivates us to propose two DNNs, dedicated to the subcarrier matching and passive beamforming, respectively.}, where one DNN is designed for subcarrier matching, and the other DNN is designed for RIS passive beamforming. 
	
	\subsubsection{DNN Structure}
	The proposed DNN for the subcarrier matching design consists of the following: There are six layers. The first layer is the input layer, where the $\{A_p, B_q\}$ for Case-I or $\{A_p, B_q, C_p\}$ for Case-II and initial RIS passive beamforming matrices are reshaped into the image and real numbers as inputs. The second, third, and fourth layer is fully-connected (FC) feed-forward layers, which includes $10NM$ neurons in each layer. The final layer is the output layer.  Besides, the activation function is leaky-Relu. The batch-normalization (BN) is adopted, and the loss function is cross-entropy (CE). The DNN for subcarrier matching is illustrated in Fig. \ref{ELF}. According to \cite{goodfellow2016deep},  the CE loss of input $x$ and output class $y$ can be expressed by
	\begin{equation}
		\text{Loss}_{\text{CE}}(x,y) = -\log \dfrac{e^{x[y]}}{\sum_{j=1}^J e^{x[j]}}, \label{crossentropy}
	\end{equation}
	where $J$ is the number of classes.
	
	The proposed DNN for RIS passive beamforming consists of the following: There are six layers. The first layer is the input layer, where $\{\textbf{v}_p,\textbf{w}_q\}$ for Case-I or $\{\textbf{v}_p,\textbf{w}_q,\textbf{C}_{p,q}\}$ for Case-II and initial RIS passive beamforming matrices are reshaped into complex and real numbers as inputs. The second, third, and fourth layer are  FC  feed-forward layers, which includes $15NM$ neurons in each layer. The final layer is the output layer, where the outputs are real numbers, mapping to the image, and real parts of complex reflecting element matrices. Besides, the activation function is leaky-Relu. The BN is adopted, and the loss function is a minimal square error (MSE). The DNN for reflecting element design is illustrated in Fig. \ref{ELF}.  According to \cite{goodfellow2016deep}, the MSE loss of input $x$ and output $y$ can be expressed by
	\begin{equation}
		\text{Loss}_{\text{MSE}}(x,y) = |x-y|^2.
	\end{equation}
	
	The complexity analysis of DNN training and inference is given below.
	
	\subsubsection{Complexity Analysis} 
	The inference and training complex of the proposed two DNNs are analyzed below. Since the number of each layer neurons of DNN for subcarrier matching is $2N + 4M$  for Case-I or $2N + N^2 + 4M$ for Case-II, $10NM$,$10NM$,$10NM$, and $N^2$, the inference complexity is thus with $\mathcal{O}(30NM + 2N + 4M + N^2)$ for Case-I and $\mathcal{O}(30NM + 2N + 4M + 2N^2)$ for Case-II. Likewise, Since the number of each layer neurons of  DNN for RIS passive beamforming $4N(M+1)^2 + 4M$ for Case-I or $4N(M+1)^2 + 2N^2M^2 + 4M$ for Case-II, $15NM$,$15NM$,$15NM$, and $4M$, the inference complexity is thus with $\mathcal{O}(45NM + 4N(M+1)^2 + 8M )$ for Case-I and $\mathcal{O}(45NM + 4N(M+1)^2 + 2N^2M^2+ 8M)$ for Case-II.  Next, for training complexity, DNN for subcarrier matching  contains $\mathcal{O}(2 \times 10^3 N^6 M^3 + 4 \times 10^3 N^5 M^4)$ for Case-I and $\mathcal{O}(2 \times 10^3 N^6 M^3 + 10^3 N^7 M^3 + 4 \times 10^3 N^5 M^4)$ parameters need to be tuned, respectively. For training complexity, The DNN for RIS passive beamforming contains  $\mathcal{O}(16 \times 15^3 (N^3M^4(M+1)^2 + N^3M^5))$ for Case-I and $\mathcal{O}(16 \times 15^3 (N^3M^4(M+1)^2 + N^3M^5) + 8 \times 15^3 N^5M^6)$ parameters need to be tuned, respectively.

	\subsection{Framework of Learning-to-Optimize Approach} 
	
	In this subsection, we present the framework of the learning-to-optimize approach in detail, which consists of the problem instance collection phase, training phase, and inference phase.
	
	\subsubsection{Problem Instance Collection Phase}
	We collect the problem instance from randomly generated wireless channels, where the distribution of wireless channels is the same. For each
	problem instance, to reduce the problem instance generation time, we employ the proposed difference-of-convex penalty-based 
	algorithm to derive the near-optimal solution.  It should be noted that in practical scenarios if the real data collection is insufficient due to a large amount of problem instance requirements, one can rely on the practical channel model, e.g., \cite{3001}, and estimate the channel statistics instead.

	\subsubsection{Training Phase}
	
	The training and testing data are generated by running the proposed algorithms. The inputs of algorithms, (i.e., CSI and initial passive beamforming matrices), are generated by random distributions. After obtaining the outputs of algorithms, we split them into two sets (i.e., the training set and the testing set). The training process of DNNs consists of a few epochs. At each epoch, we first divide the training set into several batches and feed the DNNs with each batch of data, where the optimizer is Adam. Then, the performance of current DNNs is evaluated on the testing set. This training process is repeated until the convergence of both training and testing curves.
	
	\subsubsection{Inference Phase}
	After two DNNs are trained, the model will be downloaded to the devices for inference purposes. The inference is conducted in real environments for practical use.
	
	
	\section{Simulations}
	
	\begin{figure} 
		\centering
		\includegraphics[width=2.8in]{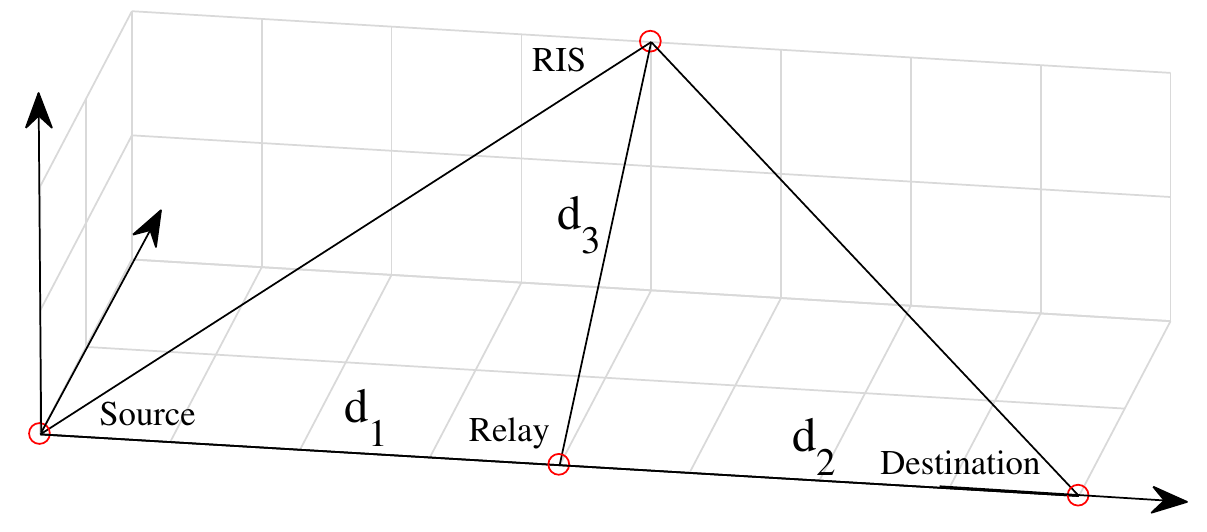}
		\caption{Illustration of the position of the source, relay, destination, and RIS.}\label{top}
		
		\centering
		\includegraphics[width=2.8in]{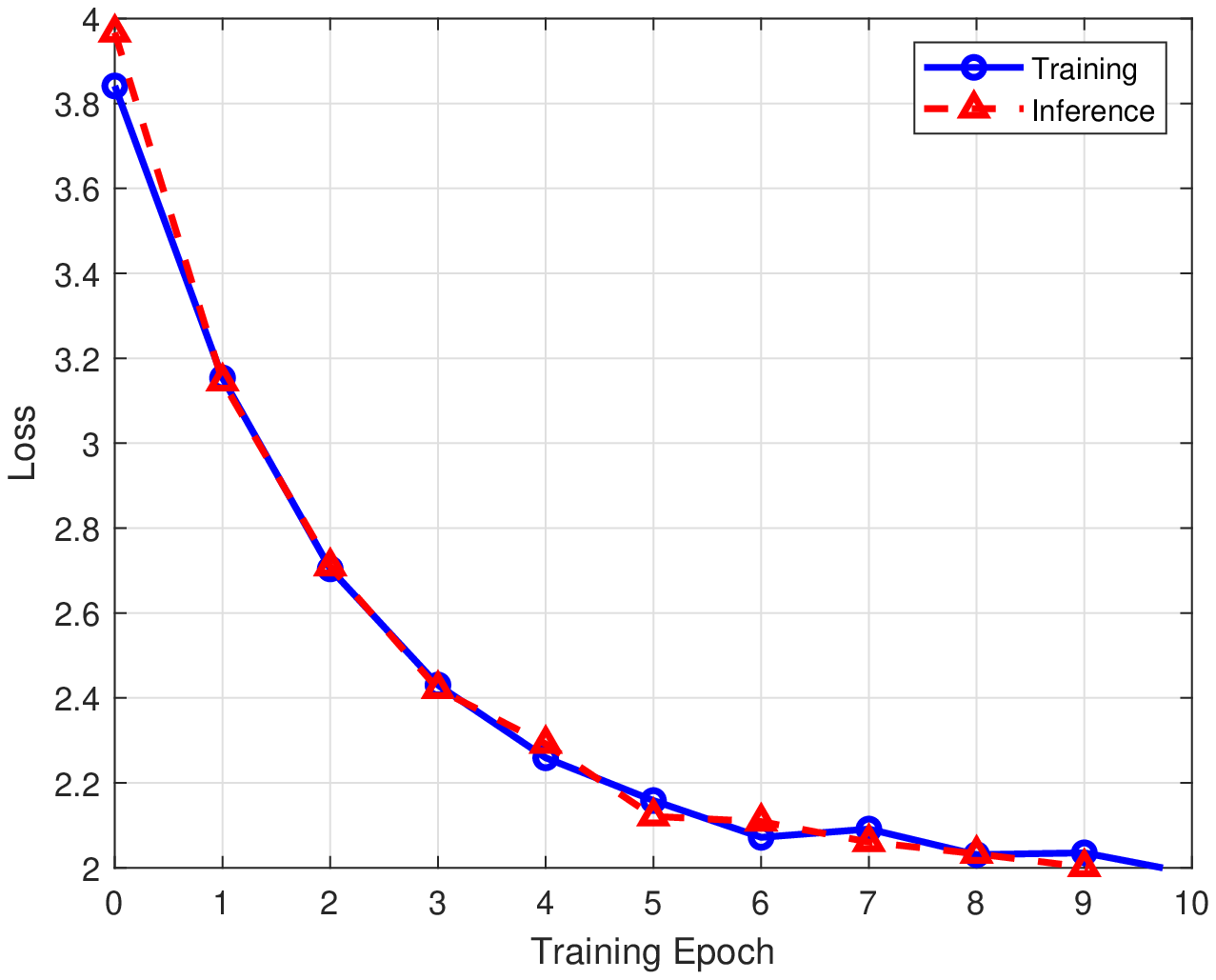}
		\caption{DNN for subcarrier matching.} \label{FL1}
		
		\centering
		\includegraphics[width=2.8in]{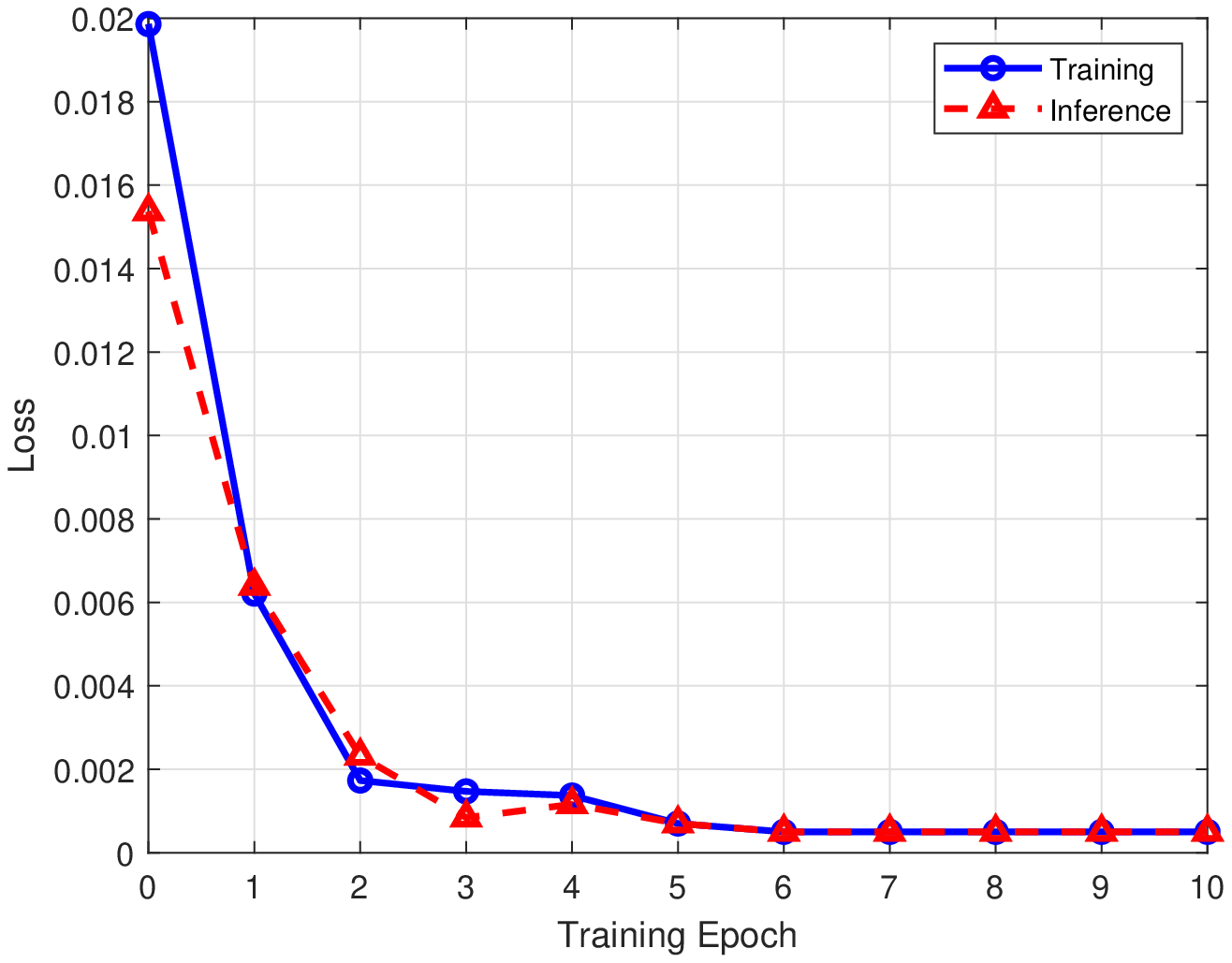}
		\caption{DNN for RIS passive beamforming.} \label{FL3}
		
	\end{figure}

	In this section, we perform computer simulations to validate the proposed algorithms and discuss the simulation results. The simulation setup is given as follows: The bandwidth of one subcarrier (i.e., $\Delta$) is given by $15$KHz. The duration of one-time slot  (i.e., $T$) is given by $100$ms. The AWGN power is given by $-90$dBm.  The cyclic prefix is sufficient to suppress the intersymbol
	interference. There are two taps, i.e., $L=2$. For each tap, it is generated from circularly symmetric complex Gaussian (CSCG) with zero mean unit variance and further degraded by large-scale fading.
	According to \cite{10}, the model of large-scale fading in dB is given by
	\begin{equation}
		L(d) = \text{PL}_1 + 10\log_{10}\left( \frac{d}{D_1} \right)^{-\alpha} + \text{Shad.},
	\end{equation}
	where $\text{PL}_1$ denotes the path-loss at the reference distance $D_1$ and is set to $-20$dB, $d$ denotes the Euclidean distance, $\alpha$ denotes the large-scale fading factor, and $\text{Shad.}$ denotes the impact of the shadowing effect. 
	The distance between source and relay is denoted by $d_1$, the distance between relay and destination is denoted by $d_2$,  and the distance between RIS and relay is denoted by $d_3$. Since the RIS can be placed on a UAV \cite{61}, we consider a RIS mounted on a UAV at the height of $1/\sqrt{2}$m if stated otherwise. The placement of source, relay, destination, and RIS is depicted in Fig. \ref{top}. The details of the simulation results and discussion are presented below.

	\subsection{Simulation Results}
	
	Before presenting the simulation results, we first introduce the abbreviations for simulated algorithms, listed as follows:
	\begin{itemize}
		\item  \textbf{BnB-I} stands for the BnB-based alternating optimization algorithm (i.e., Algorithm 2) for Case-I.
		\item   \textbf{DCP-I} stands for the difference-of-convex penalty-based algorithm (i.e., Algorithm 3) for Case-I.
		\item   \textbf{Learn-I} stands for the learning-to-optimize approach (presented in Section VI) for Case-I.
		\item  \textbf{BnB-II} stands for the BnB algorithm (i.e., Algorithm 2) for Case-II.
		\item  \textbf{DCP-II} stands for the difference-of-convex penalty-based algorithm (i.e., Algorithm 3) for Case-II.
		\item  \textbf{Learn-II} stands for the learning to optimize approach (given in Section VI) for Case-II. 
		\item \textbf{Random-I} stands for the scheme that each RIS reflecting element is uniformly generated within $[0,2\pi]$ and subcarrier matching is obtained from the BnB algorithm for Case-I.
		\item \textbf{Random-II} stands for the scheme that each RIS reflecting element is uniformly generated within $[0,2\pi]$ and subcarrier matching is obtained from the BnB algorithm for Case-II.
		\item   \textbf{RelayOnly} stands for the scenario with relay only, where the subcarrier matching is obtained from the BnB algorithm.
	\end{itemize}


	Figs. \ref{FL1} and \ref{FL3} show the convergence of learning curves of the proposed two DNNs for $M=64$, $N=4$, $d_1=d_2=8$m, $d_3=1$m,  $\alpha = 2.2$, and  $P_1 = P_2 = 2$W, where Fig. \ref{FL1} corresponds to subcarrier matching DNN, and Fig. \ref{FL3}  corresponds to beamforming DNN. We train two DNNs on Intel i5-10400F 2.90GHz, NVIDIA GForce3090. To generate a large number of samples, we use the difference-of-convex penalty-based algorithm.  Figs. \ref{FL1} and \ref{FL3} show that the test/inference performance increases with training epochs, where the numbers of training and test/inference samples are $1\times10^6$ and $2 \times 10^5$, respectively, and the batch size is $32$, and the learning rate is $10^{-3}$. In addition, the inference performance can be further enhanced by supplying more samples \cite{408}. 
	
	\begin{figure}
		\centering
		\includegraphics[width=2.8in]{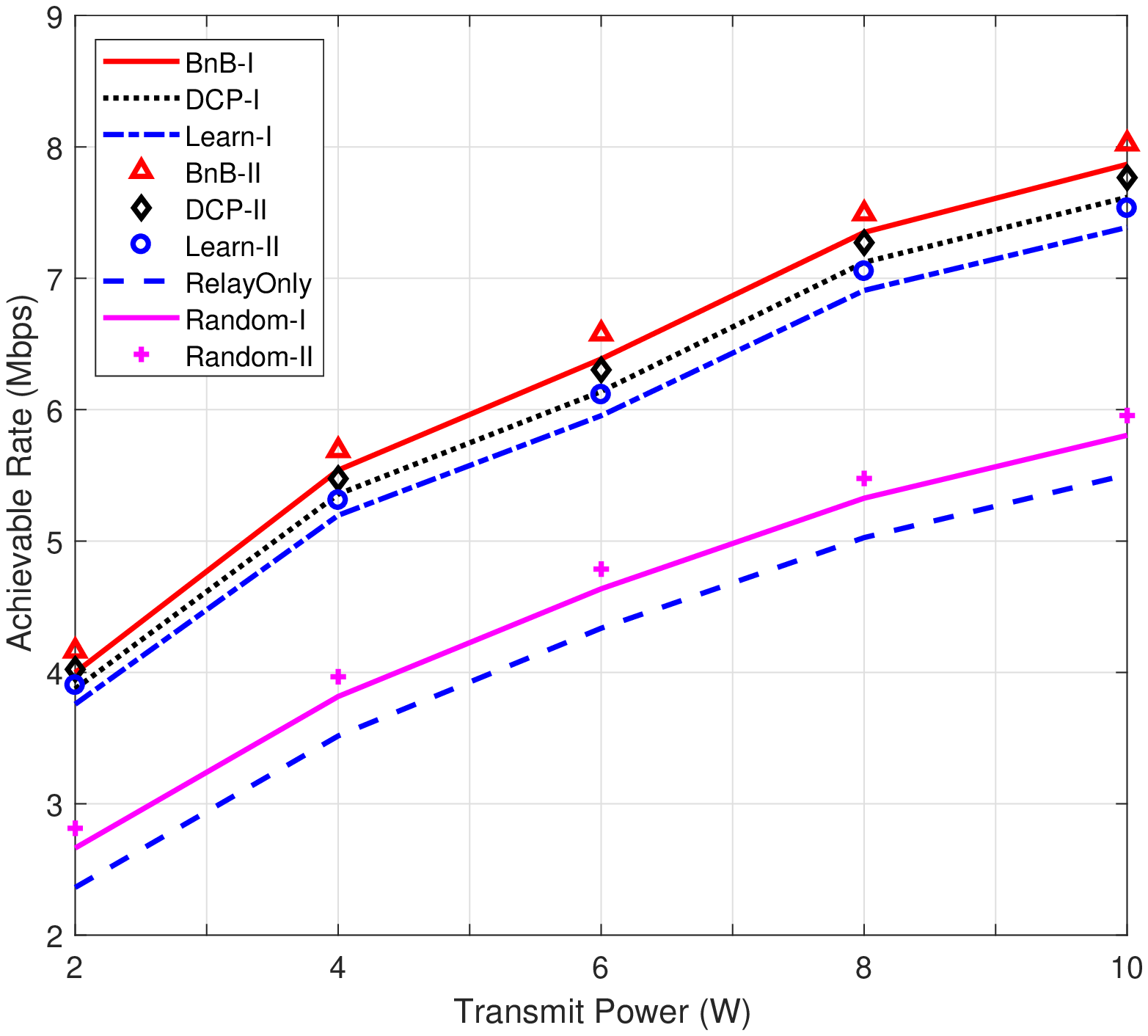}
		\caption{Achievable rate v.s. transmit power under blockage.} \label{SF1}
		\centering
		\includegraphics[width=2.8in]{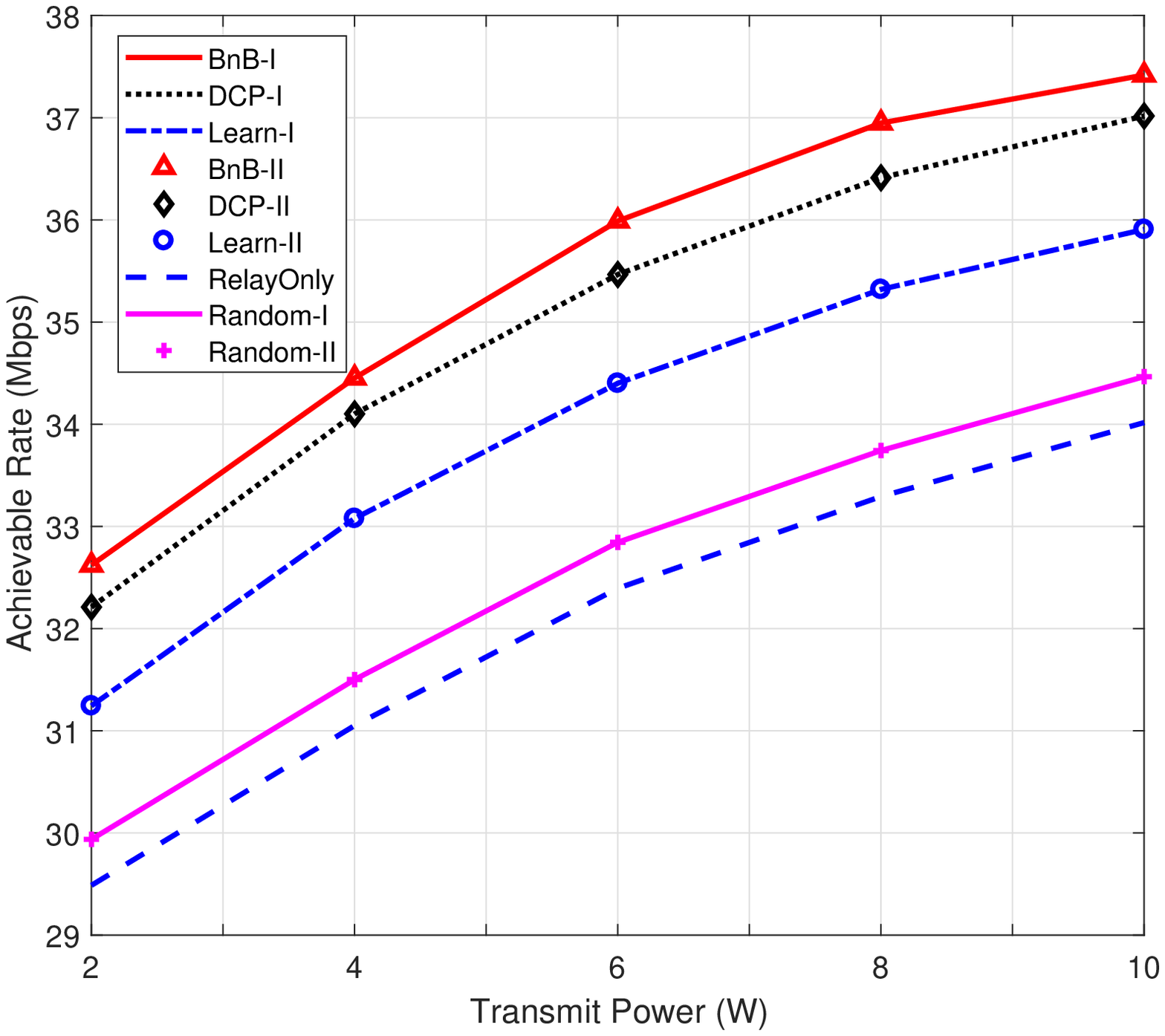}
		\caption{Achievable rate v.s. transmit power under no blockage.} \label{SF2}

	\end{figure}
	
	In Fig. \ref{SF1}, we examine the impacts of the transmit power under blockage. The simulation parameters are given by $M=64$, $N=4$, $d_1=d_2=8$m, $d_3=1$m,  $\alpha = 2.2$, and $P_1=P_2$. The links between source and relay, between relay and destination, between RIS and relay, suffer from blockage, where $\text{Shad.} = -20$dB.  Fig. \ref{SF1} shows that the achievable rate of algorithms in Case-II is higher than that in the Case-I scenario. This is due to the blockage effects in all links except for the links between source and RIS, and between RIS and destination.  Fig. \ref{SF1} shows that the performance gap between the BnB-based alternating optimization algorithm and the difference-of-convex penalty-based algorithm is marginal,  for both Cases-I and II. This shows the superior performance of the difference-of-convex penalty-based algorithm in practical settings. Moreover, the small gap between the BnB-based alternating optimization algorithm and the difference-of-convex penalty-based algorithm comes from the sub-optimality of the difference-of-convex approximation.

	\begin{figure}[t]
		\centering
		\includegraphics[width=2.8in]{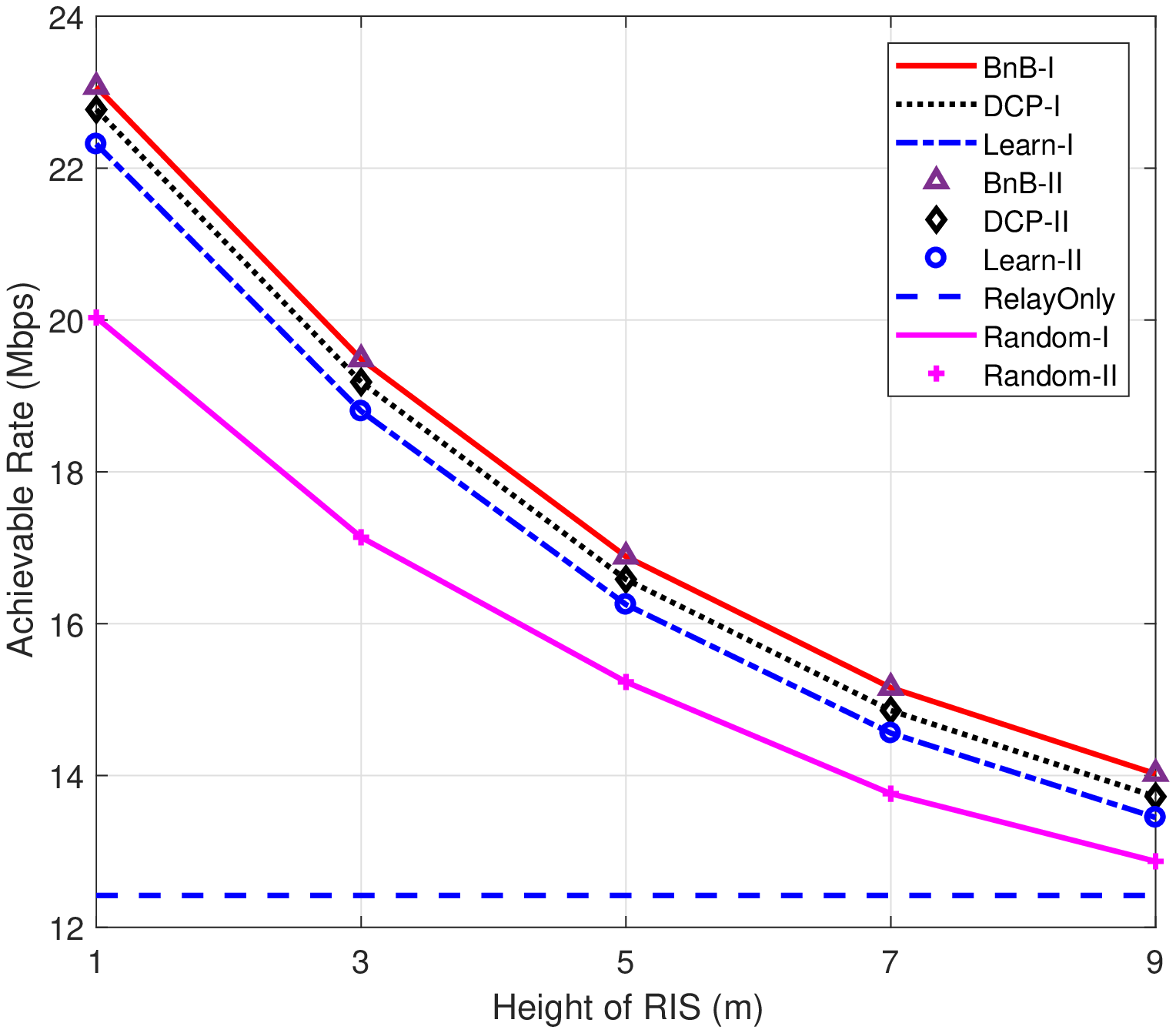}
		\caption{Achievable rate v.s. height of RIS.} \label{SF7} 
		\includegraphics[width=2.8in]{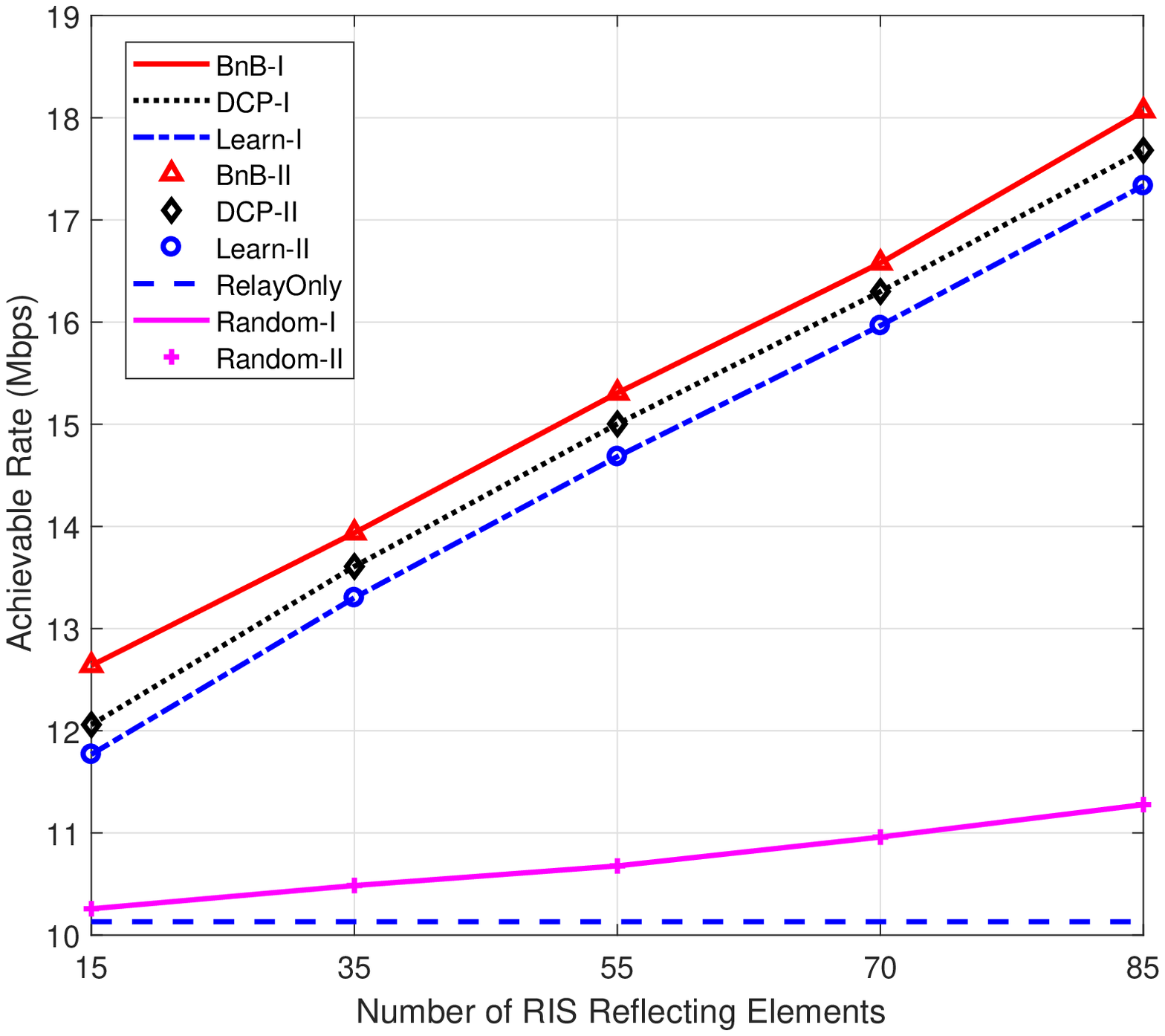}
		\caption{Achievable rate v.s. number of RIS reflecting elements.} \label{SF3} 
	\end{figure}

	In Fig. \ref{SF2}, we examine the impacts of the transmit power assuming no blockage in all considered links. The simulation parameters are the same as that in Fig. \ref{SF1}, except for $\text{Shad.} = 0$dB. Fig. \ref{SF2} shows that the achievable rates in the two cases have almost no difference. This is because the signal of RIS to destination in the first time slot suffers from ``product path-loss" (i.e., the path-loss of the source to RIS multiplies with the path-loss of RIS to destination) and the beamforming gain is mainly for the RIS to relay link, making this signal on source-RIS-destination link very weak. Compared with Fig. \ref{SF1}, when there is a blockage, Fig. \ref{SF2} shows that the achievable rate is much higher, demonstrating the severely detrimental effects brought by the blockage.

	In Fig. \ref{SF7}, by assuming that the RIS is equipped in a UAV \cite{61, Abdalla}, we examine the impact of the height of RIS. The horizontal distance from RIS to the middle of source and destination is $1$m. The remaining simulation parameters are given by $M=64$, $N=4$, $d_1=d_2=10$m,  $\alpha = 2.2$, and $P_1=P_2=1$W. Fig. \ref{SF7} shows that the achievable sum rate is inversely proportional to the height of the UAV, as the source-RIS-relay and relay-RIS-destination channels are degraded by elevating the height. Furthermore, Fig. \ref{SF7} shows that the gain of employing RIS over that of relay only system decreases as the increasing height of the UAV. 
	

	\begin{figure}
		\centering
		\includegraphics[width=2.8in]{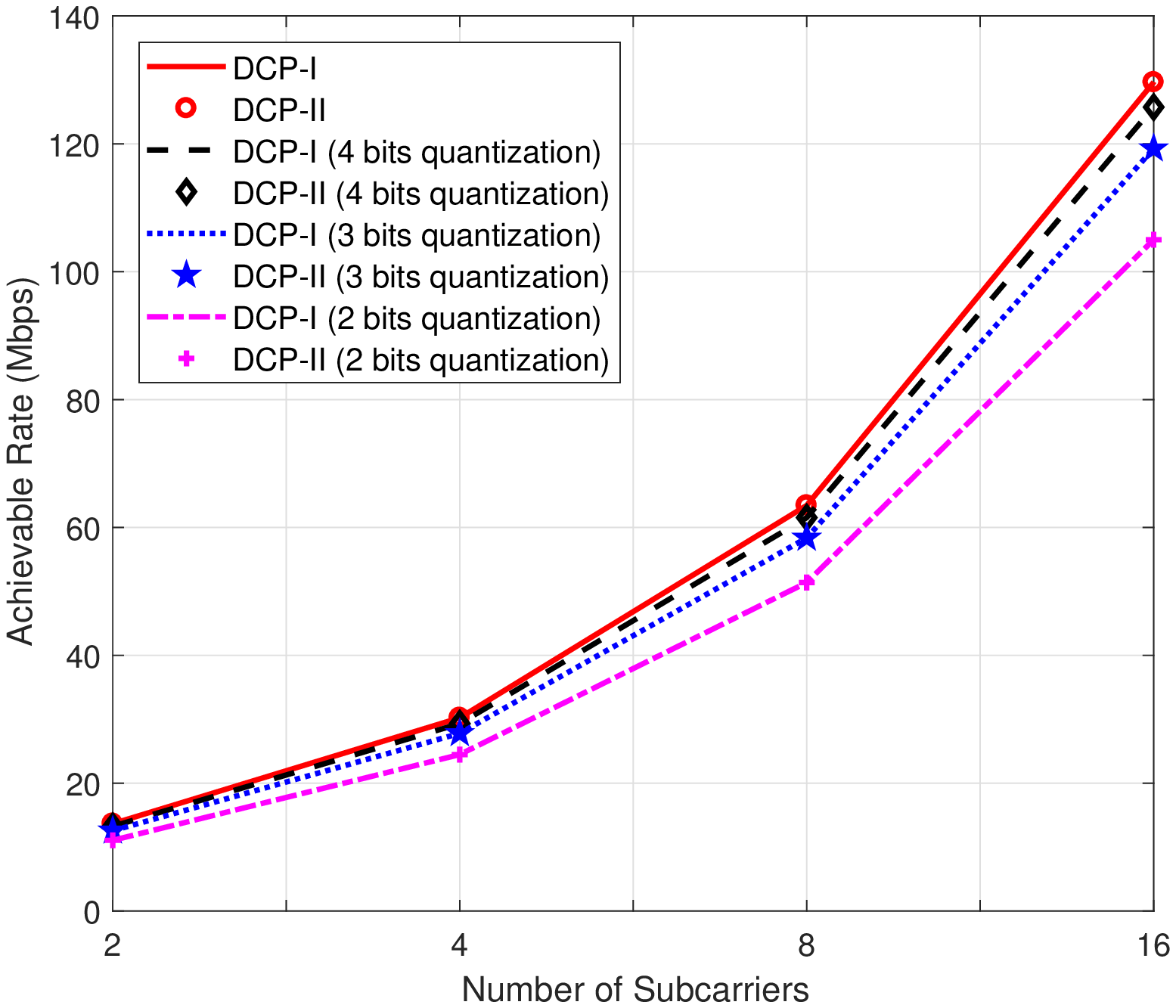}
		\caption{Achievable rate v.s. number of subcarriers.} \label{SF4}
		\centering
		\includegraphics[width=2.8in]{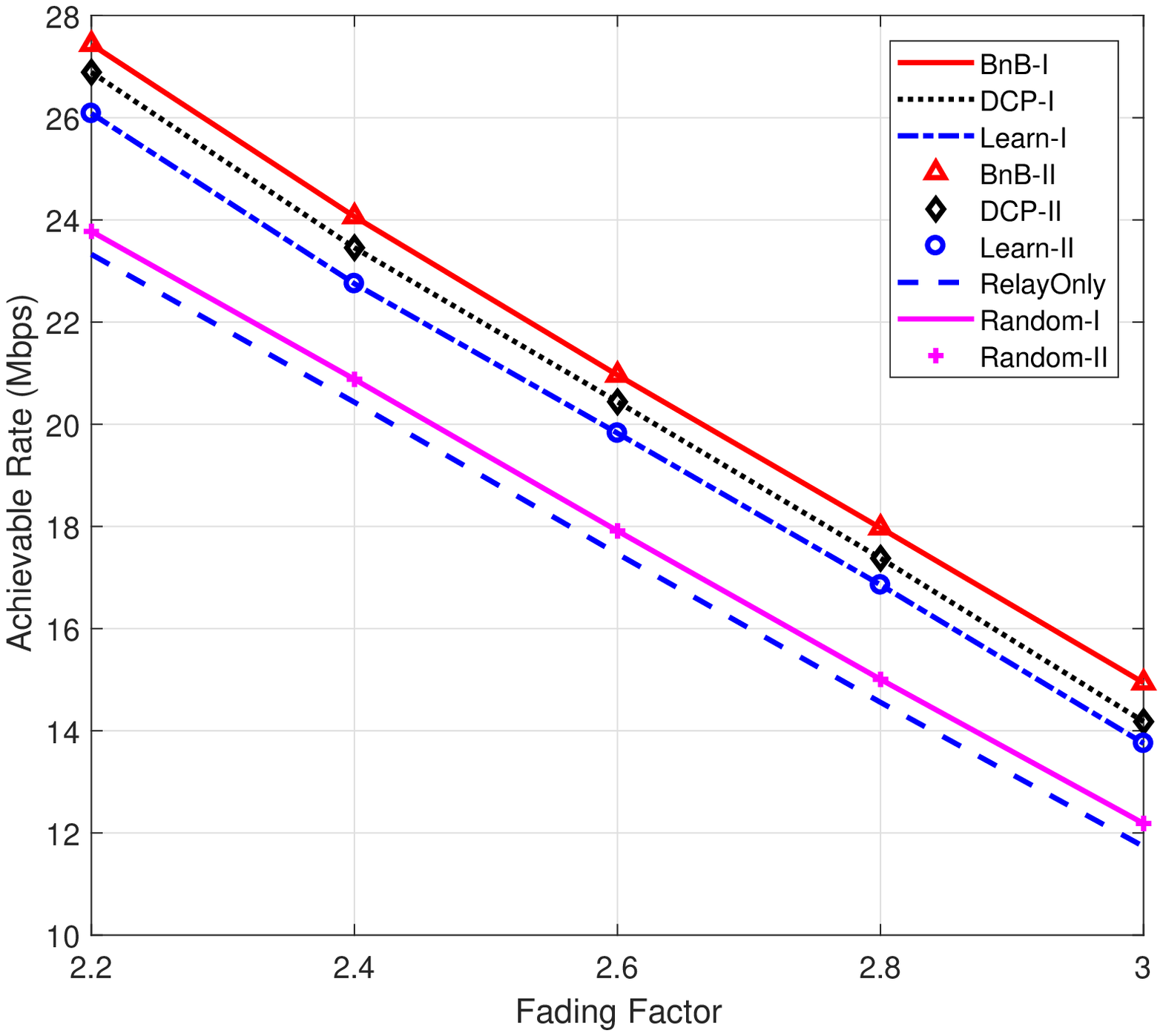}
		\caption{Achievable rate v.s. fading factor.} \label{SF8}

	\end{figure}
	In Fig. \ref{SF3}, we examine the impacts of the number of reflecting elements. The simulation parameters are given by $N=4$, $d_1=d_2=15$m, $d_3=1$m,  $\alpha = 2.2$, and $P_1=P_2=1$W.  Fig. \ref{SF3} shows that the achievable rates of the proposed algorithms increase with the number of reflecting elements. This is because, a RIS with more reflecting elements leads to a higher beamforming gain, thus improving the achievable rate. Fig. \ref{SF3} shows that the achievable rates of the proposed algorithms in the RIS-assisted relaying is much higher than that with random RIS passive beamforming and that with relay only, where the gain is magnified as the number of reflecting element increases. For OFDM relaying, this verified the power of RIS when applying it to the relay-only system and RIS-only system, coincident with that without OFDM \cite{198,199}. 
	It also shows the superior performance of the proposed optimization design of RIS passive beamforming.

	In Fig. \ref{SF4}, we examine the impacts of the number of subcarriers, and the quantization of passive beamforming. The simulation parameters are given by $M=64$, $d_1=d_2=10$m, $d_3=1$m,  $\alpha=2.2$, and $P_1=P_2=1$W. Due to the computational complexity, we only consider the difference-of-convex penalty-based algorithm. Fig. \ref{SF4} shows that the achievable rates of the proposed algorithms increase dramatically with the number of subcarriers, where the gain is magnified with the number of subcarriers. This exhibits the power of OFDM when applying it to the RIS-assisted OFDM relaying, and verifies the superiority of the proposed subcarrier matching.   Furthermore, if we round continuous phase shifts to discrete phase shifts \cite{601}, Fig. \ref{SF4} shows that the performance gap between them decreases as the number of quantization bits increases, and becomes marginal when the number of quantization bits is 4.

	In Fig. \ref{SF8}, we examine the impacts of varying fading factors. The simulation parameters are given by $M=64$, $N=4$, $d_1=d_2=10$m, $d_3=1$m,  and $P_1=P_2=1$W.  Fig. \ref{SF8} shows that the achievable rates of the proposed algorithms decrease dramatically with the fading factor.  This implies that the environment with more buildings, (i.e., a higher $\alpha$), has a much less achievable rate than that of the environment with fewer buildings, (i.e., a smaller  $\alpha$).

\begin{figure}[t] 
		\centering
		\includegraphics[width=2.75in]{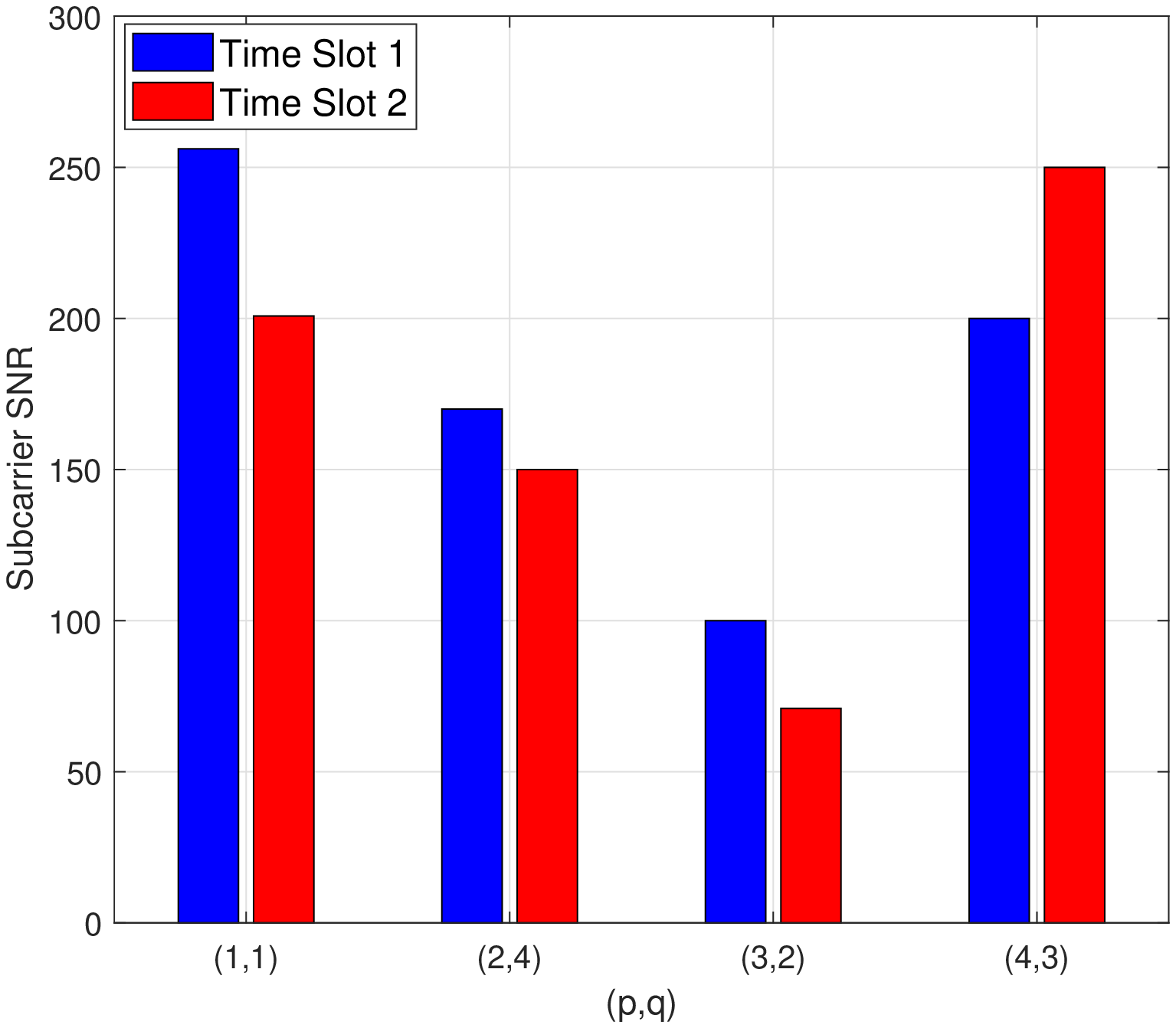}
		\caption{OFDM relaying (RelayOnly).} \label{SF5} 
		
		\centering
		\includegraphics[width=2.75in]{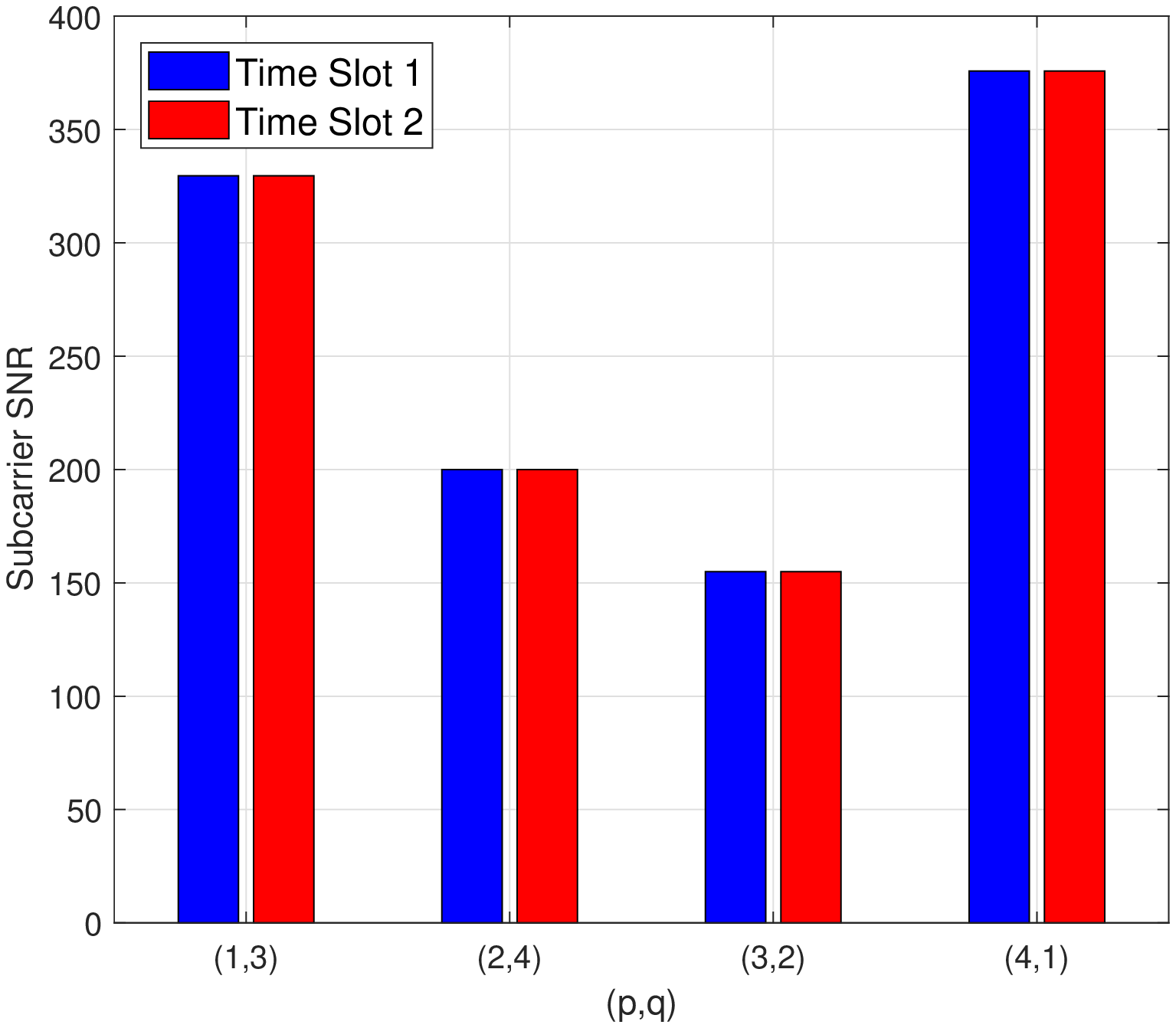}
		\caption{RIS-assisted OFDM relaying (BnBCaseI \& II).} \label{SF6}
		
		\vspace{0.3in}
		\centering
		\includegraphics[width=3in]{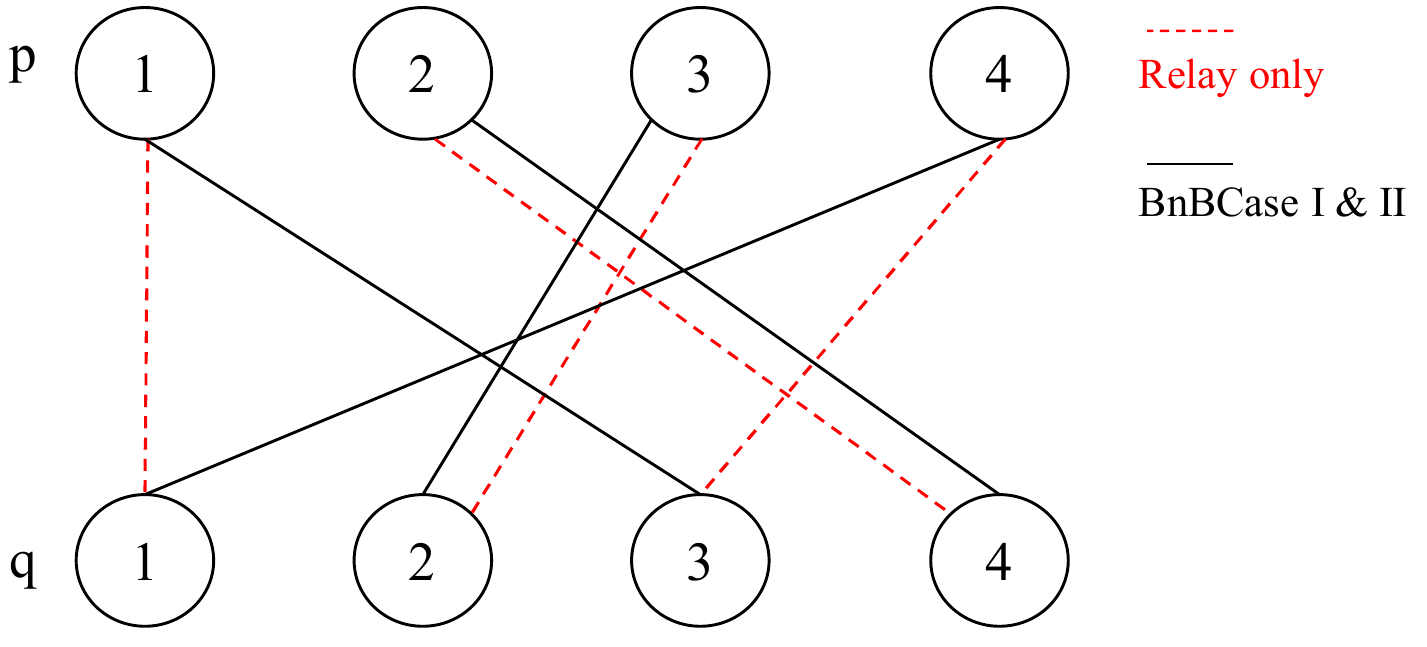}
		\caption{Comparison of subcarrier matching results.}\label{mat}
	\end{figure}

	In Figs. \ref{SF5}-\ref{mat}, we present the results on the optimized subcarrier SNRs and matching. The simulation parameters of Figs. \ref{SF5}-\ref{mat} are given by $M=64$, $N=4$, $d_1=d_2=10$m, $d_3=1$m,  $\alpha = 3$, and $P_1=P_2=1$W.  Figs. \ref{SF5}-\ref{SF6} show that compared with asymmetry matching in OFDM relaying, RIS-assisted OFDM relaying tends to balance the SNRs between the paired subcarriers. This is due to the ``$\min$" function in the objective function (see Problem $\mathcal{P}_1$).  Figs \ref{SF5}-\ref{SF6} show that RIS-assisted OFDM relaying mostly enhances the worst SNRs of matched subcarriers, to maximize the multi-carrier sum rate. Although the best strategy of matching is BTB, Fig \ref{mat} shows that the match order has been changed to satisfy the symmetry matching, in contrast to OFDM relaying.

	In Table \ref{TCom}, we examine the running time of the proposed algorithms versus the number of subcarriers, where $M=32$, $d_1=d_2=20$m, $d_3=1$m,  $\alpha=2.2$,  $P_1=P_2=1$W, and running on  Intel i5-10400F 2.90GHz. Table \ref{TCom} shows that the running time of the BnB-based alternating optimization algorithm is very high and increases dramatically with the number of subcarriers, while a significant amount of running time can be saved by using the learning-to-optimize approach. Compared with the BnB-based algorithm, the running time of the learning-to-optimize approach can be $10^{-6}$ times smaller.

	\begin{table}[t]
		\caption{Running Time Comparison} \label{TCom}
		\centering
		\begin{tabular}{|c|c|c|c|}
			\hline  
			$N$	 & BnBCaseI \& II &  DCPCaseI \& II &  LearnCaseI \& II \\ \hline
			$2$	 &  12.1192s         &  1.7442s        & \textbf{0.0876ms} \\ \hline
			$3$	 &  50.8770s        &  2.3682s        & \textbf{0.1421ms} \\ \hline
			$4$	 &  388.4235s        &  3.0205s        & \textbf{0.1947ms} \\ \hline
			$5$	 &   874.9679s     &  3.9042s        & \textbf{0.3506ms} \\ \hline
		\end{tabular}
	\end{table}
	
	\section{Conclusion}
	We considered a RIS-assisted OFDM relaying system and studied the joint design of RIS passive beamforming and subcarrier matching for sum-rate maximization in all subcarriers. The formulated problem is a MINIP, which is generally difficult to solve. To address this issue, we first proposed a BnB-based alternating optimization algorithm, which achieved a high-quality solution with exponential computational complexity. Then, we devised
	a low-complexity difference-of-convex penalty-based algorithm to reduce the complexity in BnB. To further reduce the complexity, we utilized the learning-to-optimize approach to imitate
	the proposed algorithms, which have the lowest complexity. Although these three proposed algorithms targeted the same problem, they achieved different trade-offs between sum rate and complexity, as shown by analysis and simulations. 
	Lastly, simulation results validated the effectiveness of the proposed algorithm and showed substantial performance gain of the RIS-assisted OFDM relaying system over the OFDM relaying system without RIS. In the future, it is interesting to compare the DF protocol with the amplify-and-forward (AF) protocol in RIS-assisted OFDM relaying.

	\begin{appendices}
		
		
		

		\section{Proof of Proposition 1}
		\label{AppendixB}
		
		We first denote the objective function of
		Problem $\mathcal{P}_1$ as $f({\bm{\phi}}[t],x_{p,q})$. Moreover, we assume that ${\bm{\phi}}^\text{iter}[t],x^\text{iter}_{p,q}$ are obtained by the corresponding optimization problems in the $\text{iter}^\text{th}$ iteration. Then, we have
		\begin{equation}
			f({\bm{\phi}}^{\text{iter}}[t],x^{\text{iter}}_{p,q})  \overset{(a)}{\le}  \max_{x_{p,q}} f({\bm{\phi}}^{\text{iter}}[t],x^{\text{iter}}_{p,q})   = f({\bm{\phi}}^{\text{iter}}[t],x^{\text{iter}+1}_{p,q}), \nonumber 
		\end{equation}
		where (a) is due to the optimality of the BnB algorithm. 
		In the following, we obtain
		\begin{eqnarray}
			f({\bm{\phi}}^{\text{iter}}[t],x^{\text{iter}+1}_{p,q})  \overset{(a)}{\le}  \max_{{\bm{\phi}}[t]} f({\bm{\phi}}^{\text{iter}}[t],x^{\text{iter}+1}_{p,q}) \nonumber \\ = f({\bm{\phi}}^{\text{iter}+1}[t],x^{\text{iter}+1}_{p,q}), \nonumber 
		\end{eqnarray}
		where (a) is because Problem $\mathcal{P}_5$ is convex.
		Therefore, we finally arrive at
		\begin{equation}
			f({\bm{\phi}}^{\text{iter}}[t],x^\text{iter}_{p,q}) \le f({\bm{\phi}}^{\text{iter}+1}[t],x^{\text{iter}+1}_{p,q}). \nonumber 
		\end{equation}
		That is, the objective value of Problem $\mathcal{P}_1$ is non-decreasing in
		the consecutive iterations by invoking Algorithm 2. Adding to the fact that the objective function of MINIP Problem $\mathcal{P}_1$ is upper-bounded, as this is a bounded problem. Therefore, Algorithm 2  is guaranteed to converge. 
		
		\section{Proof of Proposition 2}
		\label{AppendixC}

		In terms of worst-case computational complexity, the computing of the upper bound problem, i.e., Problem $\mathcal{P}_3$, requires a complexity of $\mathcal{O}((N^2)^{3.5})$ in worst-case by interior point method \cite{401}. There are $X_1$ times involving the computing of the upper bound problem. Moreover, there are $\mathcal{O}(2^{N^2})$ branches without pruning. While, pruning helps reduce the branches to $\mathcal{O}(X_22^{N^2})$, where $X_2$ is a parameter related to pruning. Hence, the worst-case computational complexity of BnB is $\mathcal{O}(X_1(N^2)^{3.5} + X_22^{N^2})$. It has been shown in \cite{1000} that $X_1 + X_2$ is much less than $2^{N^2}$. The worst-case computational complexity for solving Problem $\mathcal{P}_1$ given subcarrier matching by interior point method \cite{401} is $\mathcal{O}(M^{3.5})$. To sum up, the worst-case computational complexity of Algorithm 2 is $\mathcal{O}(F(M^{3.5} + X_1(N^2)^{3.5} + X_22^{N^2}))$, where $F$ is number of rounds needed for convergence.

		\section{Proof of Proposition 4}
		\label{AppendixD}
		
		Since Problem $\mathcal{P}_3$ is convex, the worst-case computational complexity of solving Problem $\mathcal{P}_3$ using interior point method is $\mathcal{O}((N^2)^{3.5})$ \cite{401}. Then, for the difference-of-convex procedure, the convergence requires $L$ rounds, where $L$ is a finite number and not very large in practice. Secondly, since Problem $\mathcal{P}_6$ is also convex, the worst-case computational complexity of solving Problem $\mathcal{P}_6$ using the interior point method is $\mathcal{O}(M^{3.5})$ \cite{401}. Finally, the convergence of alternating optimization requires $L$ rounds, where $L$ is a finite number and not very large in practice. To summarize, the worst-case computational complexity of Algorithm 3 is $\mathcal{O}(L(L(N^2)^{3.5}+M^{3.5}))$.

		\section{Gaussian Randomization}
		
		The Gaussian randomization procedure for generating rank-1 solutions is summarized as Algorithm 4.
		\begin{algorithm}[h]
			\caption{Gaussian Randomization for Rank-1 Solutions}
			\label{alg:algorithm-label}
			\begin{algorithmic}[1]
				\State \textbf{Input}: Obtaining ${\bm{\Phi}}_t,t=1,2$ from Problem $\mathcal{P}_5$
				\State Eigendecomposition: ${\bm{\Phi}}_t = \textbf{U}_t{\bf{\Sigma}}_t\textbf{U}_t',t=1,2$ 
				\State \textbf{For} $t=1:2$
				\State \quad \textbf{If} the rank of ${\bf{\Sigma}}_t$ is $1$  
				\State \quad \quad ${\bm{\Phi}}_t = \widetilde{\textbf{U}}_t(1,:)\sqrt{{\bf{\Sigma}}_t(1,1)}$
				\State \quad \textbf{Otherwise} 
				\State \quad \quad Initialize ${\cal{D}}_t$ as an empty set 
				\State \quad \quad \textbf{For} $d=1:D$ 
				\State \quad \quad Generate $\xi = \textbf{U}_t\sqrt{{\bf{\Sigma}}_t}\textbf{r}_t$ with $\textbf{r}_t$ follows ${\cal{CN}}(\textbf{0},\textbf{I}_{N+1})$
				\State \quad \quad \quad \textbf{If} $t=1$ and $\xi$ meets  SNR  constraints
				\State \quad \quad \quad \quad ${\cal{D}}_t = {\cal{D}}_t \cup  \frac{\xi}{||\xi||_2}$, where $V_{t,d}$ denotes the object-
				\State \quad \quad \quad \quad tive  value of  $\frac{\xi}{||\xi||_2}$. 
				\State \quad \quad \quad \textbf{End If}
				\State \quad \quad \textbf{End For}
				\State \quad \textbf{End If}	
				\State \textbf{End For}
				\State${\bm{\Phi}}_t = V_{t,\arg \max {{\cal{D}}_t}},t=1,2$ 
				\State  \textbf{Output}: 		$\textbf{v}_t = \exp\left(j\arg\left\{ \frac{{\bm{\Phi}}_t}{{\bm{\Phi}}_t(N+1)}\right\}_{1}^N\right),t=1,2$ 
			\end{algorithmic}
		\end{algorithm}

	\end{appendices}
	
	\bibliographystyle{IEEEtran}
	\bibliography{TVTbib}

\end{document}